\documentclass[aps,prd,nofootinbib,preprint]{revtex4}
\usepackage{amssymb}
\usepackage{amsmath,color}
\usepackage{epsfig}
\usepackage{graphicx,epsf,epsfig}
\usepackage{bbm}
\usepackage{subfigure}
\usepackage{multirow}
\usepackage{setspace}
\usepackage{ulem}
\usepackage{pifont}

\newcommand\T{\rule{0pt}{2.8ex}} % Top strut
\newcommand\Tbig{\rule{0pt}{4.5ex}} % Top strut
\newcommand\B{\rule[-1.2ex]{0pt}{0pt}} % Bottom strut
\newcommand{\ul}{\underline}
\newcommand{\mee}{\langle m_{ee}\rangle}

\newcommand{\dms}{\Delta m_{\rm S}^2}
\newcommand{\dma}{\Delta m_{\rm A}^2}

\newcommand{\sssol}{\sin^2\!\theta_{12}}
\newcommand{\ssatm}{\sin^2\!\theta_{23}}
\newcommand{\ssre}{\sin^2\!\theta_{13}}

\newcommand{\obb}{0\nu\beta\beta}

\newcommand{\ba}{\begin{array}{c}}
\newcommand{\baz}{\begin{array}{cc}}
\newcommand{\bad}{\begin{array}{ccc}}
\newcommand{\bav}{\begin{array}{cccc}}
\newcommand{\baf}{\begin{array}{ccccc}}
\newcommand{\ea}{\end{array}}

\def\be{\begin{equation}}
\def\ee{\end{equation}}
\def\gs{\mathrel{
   \rlap{\raise 0.511ex \hbox{$>$}}{\lower 0.511ex \hbox{$\sim$}}}}
\def\ls{\mathrel{
   \rlap{\raise 0.511ex \hbox{$<$}}{\lower 0.511ex \hbox{$\sim$}}}}

\newcommand{\bea}{\begin{equation} \begin{array}{c}}
\newcommand{\eea}{ \end{array} \end{equation}}
\newcommand{\D}{\displaystyle}

\def\slc#1{\setbox0=\hbox{$#1$}           % set a box for #1
    \dimen0=\wd0                                 % and get its size
    \setbox1=\hbox{/} \dimen1=\wd1               % get size of /
    \ifdim\dimen0>\dimen1                        % #1 is bigger
       \rlap{\hbox to \dimen0{\hfil/\hfil}}      % so center / in box
       #1                                        % and print #1
    \else                                        % / is bigger
       \rlap{\hbox to \dimen1{\hfil$#1$\hfil}}   % so center #1
       /                                         % and print /
    \fi}
\begin{document}
 \mbox{}\vspace{1.5cm}
%----------------------------------------------------------------------------------
\title{\Large Sterile Neutrinos for Warm Dark Matter and the Reactor Anomaly in Flavor Symmetry Models}
% \mbox{}\vspace{1cm}
%----------------------------------------------------------------------------------
%\date{\today}
%----------------------------------------------------------------------------------

\author{James Barry}
\email{james.barry@mpi-hd.mpg.de}

 \affiliation{Max-Planck-Institut f{\"u}r Kernphysik, Postfach
 103980, 69029 Heidelberg, Germany}

\author{Werner Rodejohann}
\email{werner.rodejohann@mpi-hd.mpg.de}

 \affiliation{Max-Planck-Institut f{\"u}r Kernphysik, Postfach
 103980, 69029 Heidelberg, Germany}

\author{He Zhang}
\email{he.zhang@mpi-hd.mpg.de}

\affiliation{Max-Planck-Institut f{\"u}r Kernphysik, Postfach
103980, 69029 Heidelberg, Germany}

\begin{abstract}
\noindent
We construct a flavor symmetry model based on the tetrahedral group
$A_4$ in which the right-handed neutrinos from the seesaw mechanism
can be both keV warm dark matter particles and eV-scale sterile
neutrinos. This is achieved by giving the right-handed neutrinos
appropriate charges under the same Froggatt-Nielsen  symmetry
responsible for the hierarchy of the charged lepton masses. We discuss
the effect of next-to-leading order corrections to deviate the zeroth
order tri-bimaximal mixing. Those corrections have two sources: $(i)$
higher order seesaw terms, which are important when the seesaw
particles are eV-scale, and $(ii)$ higher-dimensional effective
operators suppressed by additional powers of the cut-off scale of the
theory. Whereas the mixing angles of the active neutrinos typically
receive corrections of the same order, the mixing of the sterile
neutrinos with the active ones is rather stable as it is connected
with a hierarchy of mass scales. We also modify an effective $A_4$
model to incorporate keV-scale sterile neutrinos.
\end{abstract}
\maketitle
%%%%%%%%%%%%%%%%%%%%%%%%%%%%%%%%%%%%%%%%%%%

\newpage

\section{Introduction} \label{sect:intro}

Apart from the direct proof of physics beyond the Standard Model
(SM) in the form of neutrino masses~\cite{Nakamura:2010zzi}, a
somewhat more indirect proof is the presence of Dark Matter
(DM)~\cite{Bertone:2010zz}. One can take the point of view that
these two aspects of physics beyond the SM are connected with each
other, i.e.~that neutrino mass and Dark Matter are linked. We will
assume this connection in the present paper.

The most direct such relationship would be realized if the light
massive neutrinos whose oscillations we observe in the lab are the
DM particles. However, they would be Hot Dark Matter, and
cosmological data is compatible only with a very small component of
this form of DM, which in fact allows one to set limits on neutrino
mass~\cite{Hannestad:2010kz}. Typically the DM is assumed to be of
the Cold Dark Matter (CDM) type, for which a WIMP (weakly
interacting massive particle), as predicted in many supersymmetric
theories, is the most popular candidate. However, Warm Dark Matter
(WDM) is another possibility compatible with observations, and in
fact could solve some of the problems of the CDM paradigm, in
particular by reducing the number of Dwarf satellite galaxies or
smoothing the cusps in the DM halos. At this point one should note
that a sterile neutrino with mass at the keV scale and with small
mixing to the active neutrinos is a WDM candidate if a
mechanism~\cite{Dodelson:1993je,Shi:1998km} to generate the correct
amount of relic population is present\footnote{keV sterile neutrinos
could also provide an explanation for pulsar
kicks~\cite{Kusenko:1997sp,Fuller:2003gy}.}. See the
reviews~\cite{Boyarsky:2009ix,Kusenko:2009up,deVega:2011si} for summaries of
mechanisms and the status of keV sterile neutrinos as DM.\\

Sterile neutrinos heavier than the active ones are an ingredient of
the seesaw
mechanism~\cite{Minkowski:1977sc,Yanagida:1979as,Glashow,GellMann:1980vs,Mohapatra:1980yp},
whose existence is strongly hinted at from the fact that active
neutrino masses are extremely small. Here, however, the right-handed
neutrinos are ``naturally'' of order $10^{10}$ to $10^{15}$ GeV, and
if one wishes to make one of them a WDM candidate one has to arrange
for this mass to come down to the keV level. The following
possibilities exist:
\begin{itemize}
\item theories with extra dimensions can exponentially suppress
fermion masses, by localizing them on a distant brane, for instance.
This has been proposed to generate seesaw neutrinos of keV scale
in~\cite{Kusenko:2010ik}, see also~\cite{Adulpravitchai:2011rq};
\item flavor symmetries~\cite{Altarelli:2010gt,Ishimori:2010au} can
predict that one of the heavy neutrino masses is zero. Slightly
breaking this symmetry generates a neutrino with much smaller mass
than the other two, whose masses are allowed by the symmetry. This
has been proposed to generate seesaw neutrinos of keV scale
in~\cite{Shaposhnikov:2006nn,Lindner:2010wr}, see
also~\cite{Mohapatra:2005wk};
\item while the commonly studied flavor models with non-abelian
discrete symmetries cannot produce a non-trivial hierarchy between
fermion masses, the Froggatt-Nielsen mechanism is capable of
this~\cite{Froggatt:1978nt}. This has been proposed to generate
seesaw neutrinos of keV scale in~\cite{Merle:2011yv}, see
also~\cite{Barry:2011wb};
\item extensions or variants of the canonical type I seesaw often
contain additional mass scales, which can be arranged to generate
keV-scale particles.  This has been proposed to generate seesaw
particles of low scale in~\cite{Chun:1995js}, see also
\cite{Barry:2011wb}.
\end{itemize}
Note that both the Froggatt-Nielsen and extra-dimensional approaches
require that the three right-handed neutrinos cannot be identified
as a triplet of a flavor symmetry, which is very often the case in
flavor symmetry models (see for instance the classification table
for $A_4$ models in Ref.~\cite{Barry:2010zk}). Furthermore, in that
case there is no overall effect on the leading order seesaw formula
$M_D^2/M_i$, with $M_i$ being the right-handed neutrino mass and
$M_D$ the Dirac mass. Both mechanisms will suppress $M_i$
quadratically, while $M_D$ is linearly suppressed, and
hence their combination $M_D^2/M_i$ is left constant.\\

In this paper we will apply the Froggatt-Nielsen mechanism to bring one
of the heavy neutrinos from its ``natural'' scale down to the keV level. We will
construct an explicit flavor symmetry model based on the group
$A_4$. As in many such models, there is also a Froggatt-Nielsen
symmetry $U(1)_{\rm FN}$ to generate the observed hierarchy of the charged lepton
masses; we will use this very same $U(1)_{\rm FN}$ for creating a
WDM candidate from the heavy neutrinos.

In addition, it should be noted that when one goes from, say,
$10^{15}$ GeV = $10^{24}$ eV down to keV = $10^3$ eV, it is not a
big problem to reduce the mass by another 3 orders of magnitude. In
this way one has generated one (or more) sterile neutrino(s) of order
eV. This would be very welcome to explain long-standing issues in
particle physics, astrophysics and cosmology. Those are the apparent
neutrino flavor transitions at LSND and MiniBooNE, which together
with the ``reactor anomaly''~\cite{Mention:2011rk,Huber:2011wv}
point towards oscillations of eV-scale sterile neutrinos mixing with
strength of order 0.1 with the active ones (see
Refs.~\cite{Kopp:2011qd,Giunti:2011gz} for recent global
fits\footnote{Sometimes the result of the calibration of Gallium
solar neutrino experiments \cite{Kaether:2010ag} is interpreted as
the ``Gallium anomaly'' and is considered to be an effect of sterile
neutrinos~\cite{Giunti:2010zu}.}). In addition, several hints mildly
favoring extra radiation in the Universe have recently emerged from
precision cosmology and Big Bang
Nucleosynthesis~\cite{Cyburt:2004yc,Izotov:2010ca,Hamann:2010bk,Giusarma:2011ex}.
This could be any relativistic degree of freedom or some other New
Physics effect, but has a straightforward interpretation in terms of
additional sterile neutrino species. Although some tension between
the neutrino mass scales required by laboratory experiments and the
Hot Dark Matter limits exists within the standard $\Lambda$CDM
framework, moderate modifications could arrange for
compatibility~\cite{Hamann:2011ge}. Finally, active to sterile
oscillations have been proposed to increase the element yield in
$r$-process nucleosynthesis in core collapse supernovae (which seems
to be too low in standard calculations, see
e.g.~\cite{Beun:2006ka,Tamborra:2011is}). It is rather intriguing
that indications of the presence of eV sterile neutrinos come from
such fundamentally different probes.

We note that a particular phenomenological model, the $\nu$MSM
($\nu$ Minimal Standard Model), has been
proposed~\cite{Asaka:2005an}, in which one of the seesaw neutrinos
is keV and the other two can generate the baryon asymmetry of the
Universe either via leptogenesis (if they are heavy) or via
oscillations when they have masses below the weak scale and are
degenerate enough~\cite{Boyarsky:2009ix,Asaka:2005an}. The idea to
exploit the neutrinos of the canonical seesaw mechanism to account
for the required eV and keV particles has been discussed in
Ref.~\cite{deGouvea:2006gz}, for instance. Here we provide a
reasoning for the low-lying scales and add to the framework a flavor
symmetry that yields at leading order tri-bimaximal mixing (TBM). In
addition we modify an existing effective $A_4$ model, which does not
contain the seesaw mechanism, by adding a sterile neutrino. Again,
applying appropriate Froggatt-Nielsen charges gives the correct
charged fermion mass hierarchy and a WDM particle.

As a starting point in our seesaw models, we will leave the
Froggatt-Nielsen charges of the seesaw neutrinos free, except for
the one which is doomed to be the keV WDM particle. By properly
choosing the charges of the other two, we can make one or two to be
of eV scale, or keep both heavy (below or above the weak scale).
Different and testable phenomenology in terms of short-baseline
oscillations or neutrinoless double beta decay ($\obb$) is then
present and characteristic for each scenario. For instance, if all
neutrinos are below the momentum scale 100~MeV of double beta decay,
the effective mass on which the amplitude depends cancels exactly.
This is in contrast to the usually considered analysis of sterile
neutrinos in double beta decay~\cite{Barry:2011wb} (or our effective
$A_4$ model), in which sterile neutrinos are simply added to the
three active ones and treated as independent entities.
Interestingly, this cancellation happens pairwise in our particular
model, because the columns of the Dirac mass matrix are proportional
to the columns of the lepton mixing matrix and each of the
right-handed neutrinos is responsible for generating one light
active mass. If this right-handed neutrino is lighter than 100~MeV,
then its contribution to double beta decay cancels exactly with that
of the associated light active neutrino.

We take particular care in evaluating next-to-leading order (NLO)
corrections to the model, which lead to deviations from TBM. Two
sources for those corrections are considered. If the right-handed
neutrino mass is $\sim$~eV instead of the natural value $10^{10}$ to
$10^{15}$~GeV, then NLO seesaw
terms~\cite{Schechter:1981cv,Grimus:2000vj,Hettmansperger:2011bt}
can be important. This is because the seesaw formula goes like
$M_D^2/M_i \, (1 + M_D^2/M_i^2)$. It is easy to see that if $M_i
\simeq$ eV and if $M_D^2/M_i \simeq 0.1$ eV, $M_D$ should be around
0.3 eV, and hence the NLO seesaw term $M_D^2/M_i^2$ can generate
effects in the percent regime. Another, more commonly studied source
of NLO corrections stems from higher-dimensional operators
suppressed by additional powers of the cut-off scale of the theory.
The relative magnitude of those terms also depends on details of the
model and of the scales chosen for the neutrinos and other
particles. We show in particular that values of $U_{e3}$ compatible
with recent fits~\cite{Fogli:2011qn,Schwetz:2011zk} can be obtained
in our models. An important aspect is that the three mixing angles
of the active neutrinos typically receive corrections of the same
order, as is generically the case in flavor models. However, the mixing
of the sterile neutrinos with the active ones is rather stable as it
is defined as a hierarchy of mass scales, thus stabilizing, for
instance, the small mixing of the WDM neutrino with the active ones.\\

The remaining parts of this work are organized as follows: in
Section~\ref{sect:gen} we present some model-independent features of
the seesaw mechanism and its resulting phenomenology in the case
that one or more of the right-handed neutrinos is light.
Section~\ref{sect:seesaw} introduces a seesaw model based on the
$A_4$ flavor symmetry, in which one of the three right-handed
neutrinos acts as the WDM candidate. Various cases for the mass
scales of the other two neutrinos are discussed, phenomenological
consequences are figured out in detail, and the role of higher-order
corrections is studied. Details of the NLO terms are delegated to
the Appendix. Section~\ref{sect:efftheory} details an effective
theory with a single keV sterile neutrino added to an existing $A_4$
model. Higher-order corrections and possible deviations from the
exact TBM pattern are also discussed. We summarize and conclude in
Section~\ref{sect:summary}.

\section{Light sterile neutrinos in type I seesaw}
\label{sect:gen}

Before describing a specific model, we address the role of light
sterile neutrinos in the type~I seesaw, in particular the effect of
NLO seesaw corrections to neutrino mixing parameters as well as
phenomenological consequences of light sterile states.

\subsection{NLO seesaw corrections}

In the canonical type I seesaw mechanism, one extends the SM
particle content with three right-handed neutrinos
($\nu^c_1,\nu^c_2,\nu^c_3$) together with a Majorana mass $M_R$. The
full $6\times6$ neutrino mass matrix in the basis
($\nu_e,\nu_\mu,\nu_\tau,\nu_1^c,\nu_2^c,\nu_3^c$) reads
\begin{equation}
M_\nu^{6\times 6} = \begin{pmatrix} 0 & M_D \\ M_D^T & M_R
\end{pmatrix}, \label{eq:mnu_full}
\end{equation}
where $M_D$ denotes the Dirac mass term, and we use the LR convention
for the Lagrangian. Assuming $M_R \gg M_D$, this mass matrix can be
approximately diagonalized using a $6\times 6$ unitary matrix as
\begin{equation}
U_\nu \simeq \begin{pmatrix} 1- \frac{1}{2} BB^\dagger & B \\
-B^\dagger & 1-\frac{1}{2}B^\dagger B
\end{pmatrix}
\begin{pmatrix} V_{\nu} & 0 \\ 0 & V_R \end{pmatrix},
\label{eq:fulldiag}
\end{equation}
where $B=M_D M^{-1}_R + {\cal O}\left(M_D^3 (M_R^{-1})^3\right)$
governs the effect of higher-order seesaw
corrections~\cite{Schechter:1981cv,Grimus:2000vj,Hettmansperger:2011bt}.
The matrix $V_{\nu}$ is given by
\begin{equation}\label{eq:seesaw}
M_\nu = -M_D M_R^{-1}M_D^T = V^{}_{\nu}\, {\rm diag}(m_1,m_2,m_3) V_{\nu}^T
\end{equation}
with $m_i$ being the light neutrino masses, and $V_R$ diagonalizes
the right-handed neutrino mass matrix, i.e.~$M_R = V_R \,{\rm
diag}(M_1,M_2,M_3)V_R^T$.

In the ordinary type I seesaw framework, $M_R$ is commonly chosen to
be close to the Grand Unification scale (e.g.~$M_R \simeq 10^{14}~{\rm
GeV}$), while $M_D\simeq 100~{\rm GeV}$, so that the light neutrino
masses are suppressed at the sub-eV scale, i.e.~${\cal O}(M_D^2/M_R)
\simeq 0.1~{\rm eV}$. Therefore, the NLO seesaw corrections
governed by $B$ can be safely neglected. In models with keV-scale
sterile neutrinos these corrections are also negligible.
However, for models with right-handed neutrinos located at very low-energy scales, i.e.~at the
eV scale, $M_D \simeq 0.1~{\rm eV}$ is required in order to generate
light neutrino masses. In that limit the NLO seesaw terms are
significant, and $B\simeq 0.1$ may lead to sizable corrections to
neutrino mixing parameters. In the remaining parts of this work we
will keep the NLO seesaw corrections up to ${\cal O}(B^2)$. Note that the block-diagonalization of Eq.~\eqref{eq:mnu_full} by Eq.~\eqref{eq:fulldiag} is still approximately valid, as the remaining off-diagonal terms $B M_D^2/M_R \simeq M_D^3/M_R^2\simeq 0.001$ eV are much smaller than the ${\cal O}(1)$ eV mass difference between the active and sterile neutrinos\footnote{In our numerical calculations we did not
use any approximations, but rather numerically diagonalized the full neutrino mass matrix, obtaining results consistent with the analytical calculations.}.

\subsection{Active-active and active-sterile mixing}

In the basis where the charged lepton mass matrix is diagonal, the
light active neutrinos mix via the $3\times3$ matrix
$(1-\frac{1}{2}BB^\dagger)V_\nu$, whereas the mixing between the
active neutrino $\nu_\alpha$ ($\alpha=e,\mu,\tau$) and the sterile
neutrino $\nu_i^c$ ($i=1,2,3$) is given by
\begin{equation}
\theta_{\alpha i} \equiv [U_\nu]_{\alpha,3+ i} = [B V_R]_{\alpha i} \simeq \left[M_D(V^*_R \tilde{M}_R^{-1} V_R^\dagger)V_R\right]_{\alpha i} = \frac{[M_D
V^*_R]_{\alpha i}}{M_{i}}\, , \label{eq:as_mixing_i}
\end{equation}
where $\tilde{M}_R^{-1} = {\rm diag}(M^{-1}_1,M^{-1}_2,M^{-1}_3)$.
This illustrates that active-sterile mixing is defined as the ratio of
two scales, $M_D$ and $M_R$.
The interaction between each sterile neutrino $\nu^c_i$ and the
entire active sector is
\begin{equation}
 \theta_i^2 \equiv \sum_{\alpha = e,\mu,\tau} |\theta_{\alpha
 i}|^2\; .
\label{eq:as_mixing2}
\end{equation}
In the setup described above with
eV-scale right-handed neutrinos and $M_D \simeq 0.1~{\rm eV}$,
$\theta_i = {\cal O}(M_D/M_i) \simeq 0.1$ is obtained, which could provide an
explanation for the short-baseline anomalies. In the same way, for a
keV-scale particle and the same Dirac scale $M_D \simeq 0.1~{\rm eV}$,
one gets $\theta_i \simeq 10^{-4}$ (see the discussion in the
following subsection).

With the above notation, it is not difficult to check that the standard seesaw formula in Eq.~\eqref{eq:seesaw} can be re-expressed as
\begin{equation}\label{eq:seesaw2}
[M_\nu]_{\alpha \beta} = \left[-M_D M_R^{-1}M_D^T\right]_{\alpha \beta} = -
\sum_{i=1,2,3} \theta_{\alpha i} \theta_{\beta i} M_i \; ,
\end{equation}
indicating that each sterile neutrino makes a contribution to the
active neutrino masses of order $\theta^2_{i} M_i
$~\cite{Smirnov:2006bu}. For example, for an eV-scale sterile
neutrino $M_i \simeq {\rm eV}$ together with $\theta_i \simeq 0.3$,
the contribution to the active neutrino masses is of order
$0.1~{\rm eV}$; for a GeV-scale sterile neutrino $M_i \simeq {\rm
GeV}$ to give a contribution of the same order its corresponding
mixing angle should be $\theta_i \simeq 10^{-5} $. As a general
rule, the heavier the right-handed neutrino mass, the smaller the
active-sterile mixing.

In general the charged lepton mass matrix may not be diagonal: in
that case the total $3\times6$ lepton mixing matrix connecting the
three left-handed lepton doublets $L_\alpha = (\nu_\alpha,\alpha)^T$
($\alpha = e,\mu,\tau$) to the six neutrino mass eigenstates is
\begin{equation} \label{eq:u_def}
 U %\equiv \left(V_\ell^\dagger V , V_\ell^\dagger S \right)
\simeq \left[ V_\ell^\dagger\left(1-\frac{1}{2}BB^\dagger\right)V_\nu \, , V_\ell^\dagger B V_R \right] ,
% \simeq V_\ell^\dagger \begin{pmatrix} \left(1-\frac{1}{2}BB^\dagger\right)V_\nu & \frac{M_D V_R^*}{\widetilde{M}_R} \end{pmatrix}
\end{equation}
where $V_\ell$ is defined by $M^{}_\ell M_\ell^\dagger = V^{}_\ell\, {\rm
diag}\left(|m_e|^2,|m_\mu|^2,|m_\tau|^2\right) V_\ell^\dagger$. Note
that for $B= 0$ the standard result $ V_\ell^\dagger V^{}_\nu$ for the
$3 \times 3$ lepton mixing matrix is obtained.

\subsection{keV sterile neutrino WDM}

If one of the above-mentioned sterile neutrinos is located at the keV
scale and does not decay on cosmic time scales, it could be viewed as
a WDM candidate. In realistic sterile neutrino WDM models, a specific
mechanism for the relic production of sterile neutrinos is
required. For instance, in the Dodelson-Widrow scenario,
i.e.~production by neutrino oscillations, if one assumes that sterile
neutrino WDM with mass $M_s$ and mixing $\theta_s$
makes up all the DM in the Universe, its abundance is
given by \cite{Dodelson:1993je,Abazajian:2001nj,Abazajian:2001vt,Dolgov:2000ew,Abazajian:2005gj,Asaka:2006nq}
\begin{eqnarray}
\Omega_{\rm DM} \simeq 0.2 \left(\frac{\theta^2_s}{3\times
10^{-9}}\right) \left(\frac{M_s}{3~{\rm keV}}\right)^{1.8} \,
.\label{eq:omega_DM}
\end{eqnarray}
In this work we do not focus on a specific production mechanism of
sterile neutrino WDM, but will take Eq.~(\ref{eq:omega_DM}) as a
guideline and demand that the WDM neutrino has a mass of a few keV
and mixing of order $10^{-4}$ with the active sector. Our main focus
lies on the feasibility of accommodating sterile neutrinos in flavor
models.

It should be noticed that such a light sterile neutrino ($\nu_s$) results in a contribution $\theta^2_s M_s \simeq 10^{-5}~{\rm eV}$ to the active neutrino masses, which is much smaller than the lower bound from oscillations of ${\cal O}(10^{-2})$~eV, and hence can be safely ignored when discussing active neutrino masses and mixings. Effectively, one can
decouple $\nu_s$ in the seesaw formula, leaving only a $5\times 5$
mixing matrix  together with 2 massive right-handed neutrinos, and a
$3\times5$ mixing matrix in Eq.~\eqref{eq:u_def}. We present an
explicit model in Sect.~\ref{sect:seesaw} in order to realize such
an effective picture.

\subsection{Neutrinoless double beta decay} \label{subsect:onubb}

As already mentioned in the introduction, neutrinos with mass below
$|q| \simeq 100\,\,{\rm MeV}$ contribute to the $\obb$ process via
an effective mass defined by $\langle m_{ee} \rangle = |\sum_{i=1}^n
U^2_{ei} m_i|$, where $i$ runs over all the light neutrino mass
eigenstates.  On the contrary, for right-handed neutrinos with
masses much larger than $|q|$, their effect in $\obb$ is strongly
suppressed by the inverse of their mass. Therefore, if all the
right-handed neutrinos are light, i.e.~$M^2_i \ll q^2$, one obtains
\begin{eqnarray}\label{eq:mee_6by6}
\langle m_{ee} \rangle = \left|\sum^3_{i=1} U^2_{ei} m_i +
\sum^3_{i=1}U^2_{e,3+i} M_{i}\right| = \left[M_\nu^{6 \times 6}\right]_{ee} = 0\; ,
\end{eqnarray}
showing that the effective mass cancels exactly, since the the
($1,1$) entry of the full $6\times6$ neutrino mass matrix in
Eq.~\eqref{eq:mnu_full} is vanishing. However, this cancellation is
not realized if one of the right-handed neutrinos is very heavy,
since one should decouple this heavy neutrino in computing the
amplitude for $\obb$.

The result in Eq.~\eqref{eq:mee_6by6} holds in the general framework
of type I seesaw models. However, in certain flavor symmetric seesaw
models in which neutrino mixing is entirely determined by the Dirac
mass term, $M_D$ can be expressed
as~\cite{Chen:2009um,Choubey:2010vs}
\begin{equation}
 M_D = V_\nu \, {\rm
diag}\left(\sqrt{-m_1 M_1},\sqrt{-m_2 M_2},\sqrt{-m_3 M_3}\right)
V_R^T  \, .
\end{equation}
The active-sterile mixing in Eq.~(\ref{eq:as_mixing_i}) is now given
by $\theta_{\alpha i} = U_{\alpha,3+i}=(V_{\nu})_{\alpha i}
\sqrt{-m_i/M_i}$, which is merely a rescaling of each column of
$V_\nu$, indicating a direct connection between active and sterile
sectors. Interestingly, this implies that the above-mentioned
cancellation for light right-handed neutrinos in $\langle m_{ee}
\rangle$ occurs pairwise, since
\begin{equation}
 U_{e,3+i}^2 M_i = \left[-(V^2_\nu)_{ei} \frac{m_i}{M_i}\right] M_i = -U^2_{ei}m_i\, , \quad (i=1,2,3)\, ,
\end{equation}
neglecting terms of order $B^2$ in Eq.~\eqref{eq:u_def}. Here we
have assumed $M_\ell$ to be diagonal, but the result still holds
with non-trivial $V_\ell$, which can be factored out from both
$U_{ei}$ and $U_{e,3+i}$. Put into words, this result means that the
contribution to $\langle m_{ee} \rangle$ from the $i$-th active
neutrino is exactly cancelled by the contribution from the $i$-th
sterile neutrino. This actually simplifies the computation of
$\langle m_{ee} \rangle$ since in Eq.~\eqref{eq:mee_6by6} one only
needs to count the effects of those active neutrinos whose
corresponding sterile neutrinos are heavier than $|q|$.

\section{$A_4$ seesaw model with one \lowercase{ke}V sterile neutrino} \label{sect:seesaw}

In this section we describe an $A_4$ seesaw model with three
right-handed neutrinos: one at the keV scale and the other two  at
either the eV scale, the heavy scale ($\gs$~GeV), or both. The FN
mechanism is used to control the mass spectrum of right-handed
neutrinos and to set the charged lepton mass hierarchy; since most
$A_4$ seesaw models place right-handed neutrinos in the triplet
representation (see the classification table in
Ref.~\cite{Barry:2010zk}) one has to make non-trivial modifications
to those models in order to assign different FN charges to each
sterile neutrino\footnote{The model in Ref.~\cite{King:2006np} also
has right-handed neutrinos as singlets, but instead of the FN
mechanism a hierarchy amongst the flavons is assumed.}. Indeed, in
order to get TBM~\cite{Harrison:2002er},
\begin{eqnarray}
U_{\rm TBM} = \begin{pmatrix} \sqrt{\frac{2}{3}} &
\sqrt{\frac{1}{3}} & 0 \cr -\sqrt{\frac{1}{6}} & \sqrt{\frac{1}{3}}
&  -\sqrt{\frac{1}{2}} \cr -\sqrt{\frac{1}{6}} &  \sqrt{\frac{1}{3}}
&  \sqrt{\frac{1}{2}} \end{pmatrix}\, ,
\end{eqnarray}
at leading order with diagonal right-handed neutrinos as $A_4$
singlets, one must choose the vacuum expectation value (VEV)
alignments of the flavon fields along the directions of the columns
of the TBM matrix, similar to the method outlined in
Refs.~\cite{Antusch:2004gf,King:2006hn,King:2009ap}. The crucial
point is that each light neutrino mass eigenvalue $m_i$ is then
suppressed by only one of the heavy right-handed neutrinos $M_i$, so
that one can decouple any one of the right-handed neutrinos and
still achieve TBM with the remaining two columns, at the price of
one massless active neutrino. Since $m_2 \neq 0$, it is only viable
to decouple the neutrinos that correspond to the first or third
columns, giving normal ($m_1=0$) or inverted ($m_3=0$) ordering,
respectively. The decoupled right-handed neutrino becomes the WDM
candidate.

In what follows, we will show a concrete model example in the type I
seesaw framework, and outline various possible scenarios that differ
by the mass spectra of both active and sterile neutrinos. In each
case we demand one right-handed neutrino to be at the keV scale,
whereas the other two could be at very different scales, depending
on the chosen FN charges. Each scheme exhibits distinct
phenomenological signatures.

\subsection{Outline of the leading order model} \label{subsect:seesaw_analyt}

Here we outline the model and give general analytical results,
focussing on the decoupling of the WDM sterile neutrino.
\begin{table}[tp]
\centering \caption{Particle assignments of the $A_4$ type I seesaw
model, with three right-handed sterile neutrinos. The additional
$Z_3$ symmetry decouples the charged lepton and neutrino sectors;
the $U(1)_{\rm FN}$ charge generates the hierarchy of charged lepton
masses and regulates the mass scales of the sterile states.}
\label{table:afssmodel_a} \vspace{8pt}
\begin{tabular}{c|ccccc|ccccccc|ccc}
\hline \hline \T \B Field & $L$ & $e^c$ & $\mu^c$ & $\tau^c$ & $h_{u,d}$ & $\varphi$ & $\varphi'$ & $\varphi''$ & $\xi$ & $\xi'$ & $\xi''$ & $\Theta$ & $\nu^c_{1}$ & $\nu^c_{2}$ & $\nu^c_{3}$ \\
\hline \T $SU(2)_L$ & $2$ & $1$ & $1$ & $1$ & $2$ & $1$ & $1$ & $1$ & $1$ & $1$ & $1$ & $1$ & $1$ & $1$ & $1$ \\
$A_4$ & $\ul{3}$ & $\ul{1}$ & $\ul{1}''$ & $\ul{1}'$ & $\ul{1}$ & $\ul{3}$ & $\ul{3}$ & $\ul{3}$ & $\ul{1}$ & $\ul{1}'$ & $\ul{1}$ & $\ul{1}$ & $\ul{1}$ & $\ul{1}'$ & $\ul{1}$  \\
$Z_3$ & $\omega$ & $\omega^2$ & $\omega^2$ & $\omega^2$ & $1$ & $1$ & $\omega$ & $\omega^2$ & $\omega^2$ & $\omega$ & $1$ & $1$ &  $\omega^2$ & $\omega$ & $1$  \\
%$Z_4$ & $i$ & $1$ & $1$ & $1$ & $1$ & $i$ & $-1$ & $i$ & $-i$ & $-1$ & $1$ & $1$ & $i$ & $-1$ & $1$ \\
$U(1)_{\rm FN}$ & - & $3$ & $1$ & $0$ & - & - & - & - & - & - & - & $-1$ & $F_1$ & $F_2$ & $F_3$ \\[1mm] \hline \hline
\end{tabular}
\end{table}
Table~\ref{table:afssmodel_a} shows the particle assignments of the
$A_4$ seesaw model, with right-handed neutrinos $\nu^c_i$
($i=1,2,3$) transforming as singlets under $A_4$. Three triplet
flavons $\varphi$, $\varphi'$ and $\varphi''$ are needed to
construct the columns of $M_D$ as well as the charged lepton mass
matrix, and the singlet flavons $\xi$, $\xi'$ and $\xi''$ are
introduced in order to give masses to the right-handed neutrinos and
keep $M_R$ diagonal at leading order. The NLO terms implied by the
presence of the flavons will be discussed later. The Lagrangian
invariant under the SM gauge group and the additional $A_4 \otimes
Z_3 \otimes U(1)_{\rm FN}$ symmetry is
\begin{align}
\nonumber -{\cal L}_{\rm Y} &=  \frac{y_e}{\Lambda}\lambda^3 \left(\varphi L h_d\right) e^c
+ \frac{y_\mu}{\Lambda} \lambda\left(\varphi L
h_d\right)' \mu^c + \frac{y_\mu}{\Lambda} \left(\varphi L h_d\right)''
\tau^c  \\
&+  \frac{y_1}{\Lambda}\lambda^{F_1}(\varphi L h_u) \nu^c_1 +
\frac{y_2}{\Lambda}\lambda^{F_2} (\varphi' L h_u)'' \nu^c_2 +
\frac{y_3}{\Lambda}\lambda^{F_3} (\varphi''L h_u) \nu^c_3
\label{eq:seesaw_lag} \\
\nonumber
 &+ \frac{1}{2} \left[w_1 \lambda^{2F_1}\xi \nu^c_1 \nu^c_1 + w_2\lambda^{2F_2} \xi' \nu^c_2 \nu^c_2 + w_3 \lambda^{2F_3}\xi'' \nu^c_3 \nu^c_3
 \right] + {\rm h.c.},
\end{align}
at leading order, where the notation $(ab)'$ refers to the product
of $A_4$ triplets transforming as $\ul{1}'$, etc., and $y_\alpha$,
$y_i$ and $w_i$ are coupling constants. $\lambda \equiv \langle
\Theta\rangle/\Lambda < 1$ is the FN suppression parameter, and for
simplicity we assume $\Lambda$ to be the cutoff scale of both the
$A_4$ symmetry and the FN mechanism.

If one chooses the vacuum alignment\footnote{Note that our model
contains two Higgs doublets for the up- and down-sector,
respectively, and therefore can be accomodated within supersymmetry.
The VEV alignment could in this case be arranged by ``driving
fields''~\cite{Altarelli:2005yx}.} $\langle \varphi \rangle =
(v,0,0)$, the charged lepton mass matrix is diagonal\footnote{NLO
operators will modify the structure of $M_\ell$, introducing
non-trivial mixing in the charged lepton sector (see the
Appendix).}:
\begin{align}
M_\ell &=  \frac{v_d\, v}{\Lambda}\begin{pmatrix} y_e \lambda^{3} &
0 & 0 \\ 0 & y_\mu\lambda & 0 \\ 0 & 0 &
y_\tau
\end{pmatrix} \; ,
\label{eq:m_ell}
\end{align}
where $v_d = \langle h_d \rangle$ and the charged lepton mass
hierarchy is generated by the FN mechanism. The right-handed charged
leptons $e^c$, $\mu^c$ and $\tau^c$ carry different charges under
the $U(1)_{\rm FN}$ symmetry (cf.~Table~\ref{table:afssmodel_a}),
which leads to their observed hierarchy. We will employ the same
mechanism in the right-handed neutrino sector; for the moment the FN
charges of the right-handed sterile neutrinos are left as free
parameters, allowing us to discuss different mass spectra.

As discussed in Sect.~\ref{sect:gen}, a sterile neutrino $\nu_i^c$
with mass $M_i = {\cal O}({\rm keV})$ and mixing of order
$\theta_i^2 \simeq 10^{-8}$ will give a negligible contribution to
neutrino mass, and can thus be decoupled from the seesaw mechanism.
It is then expedient to work in a $5\times5$ basis, with the Dirac
mass matrix $M_D$ a $3\times 2$ matrix and $M_R$ a $2\times 2$
symmetric matrix. This is analogous to the minimal seesaw
model~\cite{Frampton:2002qc,Guo:2006qa} and the $\nu$MSM, in which
the lightest active neutrino is massless. The mass spectrum of
active neutrinos can either have normal ordering (NO), with $m_3 \gg
m_2 \gg m_1 \simeq 0$, or inverted ordering (IO), with $m_2 \gs m_1
\gg m_3 \simeq 0$. However, there exist different scenarios
depending on the FN charges assigned to the remaining right-handed
neutrinos. In order to keep the presentation concise we give general
analytical formulae in this subsection and discuss details specific
to the mass spectrum later on.

In our model, $\nu^c_1$ is assumed to be the WDM candidate, with a mass given by
\begin{equation}
M_1 = w_1 u \lambda^{2 F_1},
\label{eq:mass_nu1}
\end{equation}
where $u = \langle\xi\rangle$. Note here that Majorana mass terms
are doubly suppressed by the FN charge. The vacuum alignment
$\langle \varphi \rangle = (v,0,0)$ means that at leading order the
first column of the Dirac mass matrix in Eq.~\eqref{eq:mnu_full} is
$(y_1vv_u \lambda^{F_1}/\Lambda,0,0)^T$, so that
the sterile neutrino $\nu^c_1$ only mixes with the electron
neutrino\footnote{NLO terms will induce mixing between $\nu_1^c$
and $\nu_{\mu,\tau}$ (cf.~Sect.~\ref{subsect:high_order_ss}).}. From
Eqs.~\eqref{eq:as_mixing_i} and \eqref{eq:as_mixing2}, the
active-sterile mixing is
\begin{equation}
\theta_{e1} \simeq \frac{[M_D]_{e1}}{M_1} = \frac{y_1 v v_u}{w_1 u \Lambda}  \lambda^{-F_1} \; ,
\label{eq:as_mixing}
\end{equation}
so that the FN charge $F_1$ actually enhances the active-sterile
mixing,
and the contribution of the sterile neutrino $\nu_1^c$ to the lightest
neutrino mass is
\begin{equation}
 m_{1,3} = \frac{y_1^2 v^2 v_u^2}{w_1 u \Lambda^2}\, .
\label{eq:m_light_negl_2}
\end{equation}
Once we fix the scale of the various flavon VEVs,
$F_1$ is fixed by the WDM constraints [which we assume to be the ones
in Eq.~(\ref{eq:omega_DM})], and the various scenarios to be
discussed will differ only by the choice of the FN charges $F_2$ and
$F_3$, i.e.~the scale of the remaining two sterile neutrinos.

With the keV sterile neutrino $\nu^c_1$ decoupled, the seesaw
proceeds with the remaining two right-handed neutrinos, $\nu^c_2$
and $\nu^c_3$. For the NO case, we assume the triplet VEV
alignments\footnote{Ref.~\cite{King:2006np} employs a radiative
symmetry breaking mechanism in order to achieve this VEV alignment.}
\begin{equation}
\langle \varphi' \rangle = (v',v',v'), \quad \langle \varphi''
\rangle = (0,v'',-v'') \; ,
\label{eq:nh_align}
\end{equation}
which result in the following $5 \times 5$ neutrino mass matrix in the basis $(\nu_e,\nu_\mu,\nu_\tau,\nu^c_2,\nu^c_3)$:
\begin{equation}
M^{5\times 5}_\nu = \begin{pmatrix} 0 & M_D \cr M^T_D &
M_R\end{pmatrix}\, , \label{eq:mnu_5by5}
\end{equation}
with the Dirac mass matrix
\begin{equation}
M^{({\rm NO})}_D = \frac{v_u}{\Lambda} \begin{pmatrix}
y_2 v'\lambda^{F_2} & 0  \\
y_2 v'\lambda^{F_2} & -y_3 v''\lambda^{F_3}
\\   y_2 v'\lambda^{F_2} &  y_3 v''\lambda^{F_3}
\label{eq:md_nh}
\end{pmatrix}
\end{equation}
and the right-handed neutrino mass matrix
\begin{equation}
M_R = \begin{pmatrix}  w_2 u' \lambda^{2 F_2} & 0
\\  0 & w_3 u'' \lambda^{2 F_3}
\label{eq:mr_2by2}
\end{pmatrix}  ,
\end{equation}
where $u'=\langle\xi'\rangle$ and $u'' = \langle\xi''\rangle$.

The neutrino masses and flavor mixing can be obtained by the full
diagonalization of $M^{5\times 5}_\nu$, i.e.~$U_\nu^\dagger
M^{5\times 5}_\nu U_\nu^* = {\rm diag} (m_1,m_2,m_3,m_4,m_5)$, where
$m_4$ and $m_5$ denote the masses of right-handed neutrinos. Since
eV-scale sterile neutrinos may be present, one should include the NLO
seesaw terms, as motivated above. Using the formalism outlined in
Eq.~\eqref{eq:fulldiag} and
Refs.~\cite{Schechter:1981cv,Grimus:2000vj,Hettmansperger:2011bt},
and assuming real matrices for simplicity, one arrives up to order
$\epsilon_{i}^2$ at {\small \begin{align} \label{eq:U-nor}
\hspace{-.53cm}U_\nu^{({\rm NO})} &\simeq \begin{pmatrix}
\frac{2}{\sqrt{6}} & \frac{1}{\sqrt{3}} & 0 & 0 & 0 \\
-\frac{1}{\sqrt{6}} & \frac{1}{\sqrt{3}} & -\frac{1}{\sqrt{2}} & 0 &
0 \\ -\frac{1}{\sqrt{6}} & \frac{1}{\sqrt{3}} & \frac{1}{\sqrt{2}} &
0 & 0 \\ 0 & 0 & 0 & 1 & 0
\\ 0 & 0 & 0 & 0 & 1 \end{pmatrix} + \begin{pmatrix} 0 & 0 & 0 &
\epsilon_1 & 0 \\ 0 & 0 & 0 & \epsilon_1 & -\epsilon_2 \\ 0 & 0 & 0
& \epsilon_1 & \epsilon_2 \\ 0 & -\sqrt{3} \epsilon_1 & 0 & 0 & 0
\\ 0 & 0 & -\sqrt{2}\epsilon_2 & 0 & 0
\end{pmatrix} %\notag \\[1mm] &
+ \begin{pmatrix} 0 & -\frac{\sqrt{3}}{2}\epsilon_1^2 & 0 & 0 & 0 \\
0 & -\frac{\sqrt{3}}{2}\epsilon_1^2 & \frac{1}{\sqrt{2}}\epsilon_2^2
& 0 & 0 \\ 0 & -\frac{\sqrt{3}}{2}\epsilon_1^2 &
-\frac{1}{\sqrt{2}}\epsilon_2^2 & 0 & 0 \\ 0 & 0 & 0 &
-\frac{3}{2}\epsilon_1^2 & 0
\\ 0 & 0 & 0 & 0 & -\epsilon_2^2 \end{pmatrix}  ,
\end{align} }
where the expansion parameters are given by
\begin{equation}
\epsilon_1 = \frac{y_2 v' v_u}{w_2 u' \Lambda}\lambda^{-F_2} \quad
{\rm and} \quad \epsilon_2 = \frac{y_3 v'' v_u}{w_3 u'' \Lambda}
\lambda^{-F_3},
\label{eq:epsilons}
\end{equation}
in analogy to Eq.~\eqref{eq:as_mixing_i}. These parameters control
the size of active-sterile mixing and NLO corrections to neutrino
masses and mixing, and will be important in the discussions of
various scenarios in the following subsections. The neutrino mass
eigenvalues are
\begin{align}
m_1 &= 0\, , \notag \\[1mm] m_2 &= %-\frac{3  y_2^2 v'^2 v_u^2 }{w_2 u' \Lambda^2}+ \frac{9 y_2^2 v'^2 v_u^2 }{w_2 u' \Lambda^2} \epsilon^2_1
m_2^{(0)}\left(1 -3\epsilon_1^2\right)\; , \notag \\[1mm] m_3 &= %-\frac{2 y_3^2 v''^2 v_u^2  }{w_3 u'' \Lambda ^2}+\frac{4 y_3^2 v''^2 v_u^2 }{w_3 u'' \Lambda^2 }\epsilon^2_2
m_3^{(0)}\left(1-2\epsilon_2^2\right)\; , \label{eq:numassesb}  \\[1mm]
m_4 &= w_2 u' \lambda ^{2 F_2} %+\frac{3  y_2^2 v'^2 v_u^2 }{w_2 u' \Lambda^2} - \frac{9 y_2^2 v'^2 v_u^2 }{w_2 u' \Lambda^2} \epsilon^2_1
-m_2^{(0)}\left(1-3\epsilon_1^2\right) , \notag \\[1mm]
m_5 &= w_3 u' \lambda ^{2 F_3} %+ \frac{2 y_3^2 v''^2 v_u^2  }{w_3 u'' \Lambda^2} -\frac{4 y_3^2 v''^2 v_u^2 }{w_3 u'' \Lambda^2 }\epsilon^2_2
-m_3^{(0)}\left(1 - 2\epsilon_2^2\right) , \notag
\end{align}
plus higher-order terms, where
\begin{equation}
 m_2^{(0)} \equiv -\frac{3  y_2^2 v'^2 v_u^2 }{w_2 u' \Lambda^2}\, , \quad m_3^{(0)} \equiv -\frac{2 y_3^2 v''^2 v_u^2  }{w_3 u'' \Lambda ^2}\, ,
\label{eq:lo_ss_mass_nh}
\end{equation}
are the leading order seesaw terms in the NO.

For the IO case the following VEV alignments are assumed:
\begin{equation}
\langle \varphi' \rangle = (v',v',v'), \quad \langle \varphi''
\rangle = (2v'',- v'',- v'') \; .
\label{eq:ih_align}
\end{equation}
The Dirac mass matrix is modified to
\begin{equation}
M^{({\rm IO})}_D = \frac{v_u}{\Lambda} \begin{pmatrix}
y_2 v'\lambda^{F_2} & 2 y_3 v''\lambda^{F_3}  \\
y_2 v'\lambda^{F_2} & -y_3 v''\lambda^{F_3}
\\   y_2 v'\lambda^{F_2} &  -y_3 v''\lambda^{F_3}
\end{pmatrix} \; ,
\label{eq:md_ih}
\end{equation}
while the right-handed neutrino mass matrix $M_R$ remains unchanged.
In this case, the diagonalization matrix approximates (up to order
$\epsilon_i^2$) to {\small \begin{align}\label{eq:U-inv}
\hspace{-.53cm}U_\nu^{({\rm IO})} &\simeq \begin{pmatrix}
\frac{2}{\sqrt{6}} & \frac{1}{\sqrt{3}} & 0 & 0 & 0 \\
-\frac{1}{\sqrt{6}} & \frac{1}{\sqrt{3}} & -\frac{1}{\sqrt{2}} & 0 &
0 \\ -\frac{1}{\sqrt{6}} & \frac{1}{\sqrt{3}} & \frac{1}{\sqrt{2}} &
0 & 0 \\ 0 & 0 & 0 & 1 & 0 \\ 0 & 0 & 0 & 0 & 1 \end{pmatrix} +
\begin{pmatrix} 0 & 0 & 0 & \epsilon_1 & 2 \epsilon_2 \\ 0 & 0 & 0 &
\epsilon_1 & -\epsilon_2 \\ 0 & 0 & 0 & \epsilon_1 & -\epsilon_2 \\
0 & -\sqrt{3}\epsilon_1 & 0 & 0 & 0 \\ -\sqrt{6}\epsilon_2 & 0 & 0 &
0 &
0 \end{pmatrix} %\notag \\[1mm] &
+ \begin{pmatrix} -\sqrt{6}\epsilon_2^2 &
-\frac{\sqrt{3}}{2}\epsilon_1^2 & 0 & 0 & 0 \\
\sqrt{\frac{3}{2}}\epsilon_2^2 & -\frac{\sqrt{3}}{2}\epsilon_1^2 & 0
& 0 & 0 \\ \sqrt{\frac{3}{2}}\epsilon_2^2 &
-\frac{\sqrt{3}}{2}\epsilon_1^2 & 0 & 0 & 0 \\ 0 & 0 & 0 &
-\frac{3}{2}\epsilon_1^2 & 0 \\ 0 & 0 & 0 & 0 & -3\epsilon_2^2
\end{pmatrix}
\end{align} }
and the neutrino masses are given by
\begin{align}
m_1 &= %-\frac{6 y_3^2 v''^2 v_u^2 }{w_3 u'' \Lambda^2}+ \frac{36 y_3^2 v''^2 v_u^2 }{w_3 u'' \Lambda^2} \epsilon^2_2
m_1^{(0)}\left(1 -6\epsilon_2^2\right) ,\notag \\[1mm] m_2 &= %-\frac{3 y_2^2 v'^2 v_u^2  }{w_2 u' \Lambda^2}+ \frac{9 y_2^2 v'^2 v_u^2 }{w_2 u' \Lambda^2 }\epsilon^2_1
m_2^{(0)}\left(1 -3\epsilon_1^2\right)  , \notag \\[1mm] m_3 &= 0\, , \label{eq:numassesb-inv}\\[1mm]
m_4 &= w_2 u' \lambda ^{2 F_2} %+\frac{3 y_2^2 v'^2 v_u^2  }{w_2 u' \Lambda^2} - \frac{9 y_2^2 v'^2 v_u^2 }{w_2 u' \Lambda^2 }\epsilon^2_1
-m_2^{(0)}\left(1-3\epsilon_1^2\right) , \notag \\[1mm] m_5 &= w_3 u'' \lambda ^{2 F_3} %+ \frac{6 y_3^2 v''^2 v_u^2 }{w_3 u'' \Lambda^2} - \frac{36 y_3^2 v''^2 v_u^2 }{w_3 u'' \Lambda^2} \epsilon^2_2
-m_1^{(0)}\left(1-6\epsilon_2^2\right) , \notag
\end{align}
where
\begin{equation}
 m_1^{(0)} \equiv -\frac{6 y_3^2 v''^2 v_u^2 }{w_3 u'' \Lambda^2}
\label{eq:lo_ss_mass_ih}
\end{equation}
is the leading order expression for the lightest mass in the IO and
$m_2^{(0)}$ is defined in Eq.~\eqref{eq:lo_ss_mass_nh}. Note from
Eqs.~\eqref{eq:U-nor} and \eqref{eq:U-inv} that the mixing pattern
$|U_{e3}|=0$ and $|U_{\mu 3}| = |U_{\tau 3}|$ is stable with respect
to higher order seesaw terms, which is
actually true to all orders in $\epsilon_i$ \cite{Hettmansperger:2011bt}.

One salient feature of the above seesaw model can be seen from
Eqs.~\eqref{eq:lo_ss_mass_nh} and \eqref{eq:lo_ss_mass_ih}: the
leading order contributions to the active neutrino masses do not
depend on the FN charges assigned to the right-handed neutrinos. The
leading seesaw mass term is $M_D^2/M_i$, so that the one unit of FN
charge $\lambda^{F_i}$ from $M_D$ cancels with the two units
$\lambda^{2F_i}$ from $M_i$. On the other hand, the NLO term
$M_D^4/M_i^3 \propto \epsilon_i^2$ does depend on the FN charge,
which therefore controls the magnitudes of NLO corrections. The
larger the charge $F_i$ (equivalent to a smaller sterile neutrino
mass), the larger the correction parameters $\epsilon_i$ become, and
thus the larger the corrections to the leading order seesaw masses.

In addition to NLO seesaw terms, one would expect higher-dimensional
operators to modify the leading order predictions of the model,
which has so far been constructed from the leading order Lagrangian
in Eq.~\eqref{eq:seesaw_lag}. The magnitude of those corrections
depends largely on the actual numerical values chosen in the model,
since they are suppressed by additional powers of the cutoff scale
$\Lambda$. Our choice of mass scales is guided by the leading order
predictions: we need $(i)$~the sterile neutrino mass and mixing to
satisfy Eq.~\eqref{eq:omega_DM}, $(ii)$~the correct scale of active
neutrino masses and $(iii)$~Yukawa couplings to be $\leq {\cal
O}(1)$. In what regards the keV sterile neutrino [see
Eqs.~\eqref{eq:mass_nu1} and \eqref{eq:as_mixing}], a rough
numerical estimate shows that with the mass scales
\begin{equation}
v \simeq 10^{11}\ {\rm GeV}, \quad u \simeq 10^{12}\ {\rm GeV}, \quad
\Lambda \simeq 10^{13}\ {\rm GeV} \; , \label{eq:ss_massscales_1}
\end{equation}
the Higgs VEV $v_u = \langle h_u \rangle \simeq  174~{\rm GeV}$ and
$\lambda \simeq 0.1$, one needs the FN charge
\begin{equation}
F_1 = 9
\end{equation}
to obtain a sterile neutrino of mass $M_1 \simeq 1$~keV with the
desired mixing angle $\theta_1^2 \simeq 10^{-8}$, with $y_1,w_1 \leq
{\cal O}(1)$. In order to stabilize the active neutrino masses
around the sub-eV scale, one can choose [together with the numbers
in Eq.~\eqref{eq:ss_massscales_1}] the scales
\begin{equation}
v'\simeq v'' \simeq  u' \simeq u'' \simeq 10^{11}\ {\rm GeV}
\label{eq:ss_massscales_2}
\end{equation}
for the flavon VEVs, and the mass splitting among active neutrinos
can be achieved by properly choosing the corresponding Yukawa
couplings, i.e.~$y_i$ and $w_i$ ($i=2,3$). For definiteness we fix
the scales of the VEVs from here on, and obtain all numerical results
using those values.

\subsection{Mixing corrections from higher-order terms} \label{subsect:high_order_ss}

As we have already mentioned, the presence of gauge singlet flavons in
the model will inevitably induce NLO corrections, which may modify the
leading order picture and affect both active and active-sterile
neutrino mixing. Indeed, modifications to TBM are required in order to
explain the T2K result that suggests non-zero $\theta_{13}$ \cite{Abe:2011sj}.
We concentrate on the effects of adding higher-order
operators to the Lagrangian in Eq.~\eqref{eq:seesaw_lag}; one could
also introduce corrections by perturbing the $A_4$ triplet VEV
alignments \cite{Honda:2008rs,Barry:2010zk}. Note that getting
non-zero $\theta_{13}$ in models designed to predict TBM is a more
general problem, and other solutions have been proposed, e.g.~in Refs.~\cite{Boudjemaa:2008jf,Goswami:2009yy,King:2011zj}.

Since the charged lepton and right-handed neutrino mass matrices are
diagonal at leading order, TBM comes solely from the structure of the
Dirac mass matrix. Without performing a detailed numerical analysis,
one can show that the higher-order corrections affect all three mass matrices: $M_\ell$, $M_D$ and $M_R$. The impact of
those corrections is controlled by the ratios of flavon VEVs to the
cut-off scale, in our case
\begin{equation}
r_1 \equiv \frac{u}{\Lambda} \simeq 0.1 \quad {\rm and} \quad r_2
\equiv \frac{u'}{\Lambda} \simeq \frac{u''}{\Lambda} \simeq
\frac{v}{\Lambda} \simeq \frac{v''}{\Lambda} \simeq 0.01\ .
\label{eq:VEV_ratios}
\end{equation}
The terms containing the VEV $\langle\xi\rangle=u=r_1 \Lambda$ have
the largest effect, and will be included in our analysis (see the Appendix); terms
containing the VEVs $u'$, $u''$, $v$, $v'$ and $v''$ are all of
relative order $r_2 \simeq 0.01$ and can be safely
neglected. Importantly for our model, the correction terms turn out to
have a negligible effect on the keV sterile neutrino mass, as well as
its mixing with the active sector. Explicitly, from Eqs.~\eqref{eq:mrevs_nlo} and \eqref{eq:md_nlo}, the corrected active-sterile mixing is
\begin{equation}
 {\theta'_{e1}}^{({\rm NO})} \simeq \theta_{e1}\left(1 +\frac{y'_1v'}{y_1v}r_1\right)%-\frac{{w'_1}^2}{w_1^2}r_1^2\right) 
\quad {\rm and} \quad {\theta'_{e1}}^{({\rm IO})} \simeq \theta_{e1}\left[1 +\left(\frac{y'_1v'}{y_1v}+2\frac{y_3v''}{y_1 v}\frac{w'_1}{w_1}\right)r_1\right] %-\frac{{w'_1}^2}{w_1^2}r_1^2\right) 
, \label{eq:theta_e1_nlo}
\end{equation}
where the dimensionless couplings $y'_1$ and $w'_1$ are defined in Eqs.~\eqref{eq:nlo_md_terms} and \eqref{eq:mr_nlo_terms}, respectively, and the leading order expression for $\theta_{e1}$ is given in Eq.~\eqref{eq:as_mixing}. In addition,
the mixing angles $\theta_{\mu 1}$ and $\theta_{\tau_1}$ become non-zero, but of the same magnitude as
$\theta_{e1}$, i.e.
\begin{align}
 {\theta'_{\mu,\tau 1}}^{({\rm NO})} \simeq \theta_{e1}\left(\frac{y'_1v'}{y_1v} \mp \frac{y_3v''}{y_1 v}\frac{w'_1}{w_1} \right)r_1 \quad {\rm and} \quad {\theta'_{\mu,\tau 1}}^{({\rm IO})} \simeq \theta_{e1}\left(\frac{y'_1v'}{y_1v} - \frac{y_3v''}{y_1 v}\frac{w'_1}{w_1} \right)r_1 \, .
\label{eq:theta_mutau1_nlo}
\end{align}
This shows that the active-sterile mixing is
stable, illustrating the point that unlike active neutrino mixing it
is defined as the ratio of two large scales, so that small changes
in $M_D$ and $M_R$ will have little effect on $\theta_{\alpha i}$
(we assume that $|w'_1| \ls |w_1|$). The WDM particle remains
decoupled from the seesaw and one can still work in the $5\times5$
basis. We show the resulting mixing matrix elements here and provide
details of the diagonalization procedure and modified neutrino mass
eigenvalues in the Appendix.

The final lepton mixing matrix is a $3\times5$ matrix connecting the
three flavors of lepton doublets to the five neutrino mass eigenstates,
and corrections from the charged lepton sector [Eq.~\eqref{eq:vlep}]
and the neutrino sector [Eq.~\eqref{eq:u_nu_nlo}] can be combined
via Eq.~\eqref{eq:u_def} to give the approximate mixing matrix
elements
\begin{align}
|U_{e3}|^2 &\simeq
\frac{r_1^2}{2}\left[\left(\frac{y'_\mu}{y_\mu}-\frac{y'_\tau}{y_\tau}\right)^2\right]
+ \frac{1}{2}(\chi-\rho_3)^2 -(\chi-\rho_3)
r_1\left(\frac{y'_\mu}{y_\mu}-\frac{y'_\tau}{y_\tau}\right)  ,
\nonumber \\
|U_{e2}|^2 &\simeq \frac{1}{3}\left[1-3\epsilon_1^2 -2\rho_2-
2r_1\left(\frac{y'_\mu}{y_\mu}+\frac{y'_\tau}{y_\tau}\right)\right]
, \notag \\
|U_{\mu 3}|^2 &\simeq
\frac{1}{2}\left[1-2\epsilon_2^2+2\frac{y'_\tau}{y_\tau}r_1+\frac{2}{3}\sigma^N_+
R\right]  , \label{eq:nlo_final_nh} \\
|U_{e,\mu 4}|^2 &\simeq \epsilon_1^2\left[1\mp2\rho_2\mp2r_1\left(\frac{y'_\mu}{y_\mu}\pm\frac{y'_\tau}{y_\tau}\right)\right] \; , \notag \\
|U_{e5}|^2 &\simeq \epsilon_2^2 \left[r_1^2\left(\frac{y'_\mu}{y_\mu}-\frac{y'_\tau}{y_\tau}\right)^2-2r_1\left(\frac{y'_\mu}{y_\mu}-\frac{y'_\tau}{y_\tau}\right)(\chi-\rho_3)+(\chi-\rho_3)^2\right] \; , \notag \\
|U_{\mu 5}|^2 &\simeq \epsilon_2^2\left(1+2r_1\frac{y'_\tau}{y_\tau}\right)\; , \notag
\end{align}
in the NO. Here the $\epsilon_i$ are generated by
NLO seesaw terms, $y'_{\mu,\tau}$ stem from corrections to the
charged lepton mass matrix, while the other parameters come from
corrections to $M_D$ and $M_R$. For the inverted
ordering we find
\begin{align}
|U_{e3}|^2 &\simeq \frac{r_1^2}{2}\left(\frac{y'_\mu}{y_\mu}-\frac{y'_\tau}{y_\tau}\right)^2 -\rho_2 r_1\left(\frac{y'_\mu}{y_\mu}-\frac{y'_\tau}{y_\tau}\right) +\frac{\rho_2^2}{2}  , \notag \\
|U_{e2}|^2 &\simeq \frac{1}{3}\left[1-3\epsilon_1^2-2\rho_2-2r_1\left(\frac{y'_\mu}{y_\mu}+\frac{y'_\tau}{y_\tau}\right)-\frac{2}{3}\sigma^I_+ G\right]   , \notag \\
|U_{\mu 3}|^2 &\simeq \frac{1}{2}\left[1+2\rho_2+2\frac{y'_\tau}{y_\tau}r_1\right]  , \label{eq:nlo_final_ih} \\%\left(\frac{y'_\tau}{y_\tau}+\frac{y'_\mu y_\mu}{y_\tau^2}\right)
|U_{e,\mu 4}|^2 &\simeq \epsilon_1^2\left[1-2\rho_2\mp2r_1\left(\frac{y'_\mu}{y_\mu}\pm\frac{y'_\tau}{y_\tau}\right)\right]  , \notag \\
|U_{e5}|^2 &\simeq 4\epsilon_2^2\left[1+r_1\left(\frac{y'_\mu}{y_\mu}+\frac{y'_\tau}{y_\tau}\right)-(\chi-\rho_3)\right]  , \notag \\
|U_{\mu 5}|^2 &\simeq \epsilon_2^2\left[1-2r_1\left(2\frac{y'_\mu}{y_\mu}+\frac{y'_\tau}{y_\tau}\right)\right] ,\notag
\end{align}
with the parameters
\begin{align}
\sigma^N_{\pm} &\equiv \chi \pm \rho_2 - \rho_3 \, ,\quad \sigma^I_{\pm} \equiv \chi \pm 3\rho_2 - \rho_3\, , \notag \\
\chi &\equiv \frac{y_1v}{y_3v''}\frac{w'_1}{w_1}r_1\, , \quad \rho_2 \equiv \frac{y'_2v''}{y_2v'}r_1\, , \quad \rho_3 \equiv \frac{y'_3v}{y_3v''}r_1\, ,\notag \\
 R &\equiv \frac{m^{(0)}_2}{m^{(0)}_3} \simeq \sqrt{\frac{\dms}{\dma}} = {\cal O}(10^{-1})\, , \label{eq:param_defs}\\
 G &\equiv \frac{m^{(0)}_1}{m^{(0)}_2-m^{(0)}_1} \simeq \frac{2\dma}{\dms} \simeq \frac{2}{R^2} = {\cal O}(10^{2})\, , \notag
\end{align}
controlling the size of the mixing terms, where $\dms$ and $\dma$
are the solar and atmospheric mass squared differences,
respectively. The dimensionless couplings $y'_{2,3}$ are defined in Eq.~\eqref{eq:nlo_md_terms}. $R$ and $G$ contain the leading order neutrino masses
from Eqs.~\eqref{eq:lo_ss_mass_nh} and \eqref{eq:lo_ss_mass_ih}:
while $R$ is quite small, $G$ is large, which is a consequence of
the two relatively large but nearly degenerate neutrino masses in
the IO, $m_1^{(0)} \simeq m_2^{(0)} \simeq 0.05$~eV. We have
expanded to first order in $R$, but $G$ remains an exact expression
in the mixing matrix. Thus in the IO we need $\sigma^I_+=\chi + 3\rho_2 - \rho_3$ to be ${\cal
O}(10^{-3})$ in order to keep the corrections to $|U_{e2}|^2$ under
control, which in turn puts a constraint on the Yukawa couplings $y'_{2,3}$ and 
$w'_1$ in Eqs.~\eqref{eq:nlo_md_terms} and \eqref{eq:mr_nlo_terms}. With $r_1 \simeq 0.1$ and
$v\simeq v'\simeq v''$, we have $\rho_{2,3} \simeq 0.1\frac{y'_{2,3}}{y_{2,3}}$ and $\chi \simeq
0.1\frac{y_1}{y_3}\frac{w'_1}{w_1}$, so that we need to assume that
$y'_{2,3} \simeq 0.01 y_{2,3} $ and $y_1 w'_1 \simeq 0.01 y_3 w_1$ in the inverted ordering. The full
neutrino mass eigenvalues are given in Eqs.~\eqref{eq:numasses_ho}
and \eqref{eq:numasses_ho_ih}: despite the appearance of $G^2$ terms
in the IO mass eigenvalues they will always be suppressed by
$(\sigma^I_+)^2$, which is constrained to be small from the mixing matrix
element $U_{e2}$.

As expected, by setting $y'_2$, $y'_3$, $w'_1$, $y'_\mu$ and $y'_\tau$ to zero in
Eqs.~\eqref{eq:nlo_final_nh} and \eqref{eq:nlo_final_ih} one
recovers the matrix elements in Eqs.~\eqref{eq:U-nor} and
\eqref{eq:U-inv}. Note that without the higher-order correction
terms $U_{e3}$ remains exactly zero, to all orders in $\epsilon_i$.
The active-sterile mixing ($U_{\alpha 4,5}$) is always proportional
to $\epsilon_i$, or in other words to a ratio of scales
[cf.~Eqs.~\eqref{eq:as_mixing_i} and \eqref{eq:epsilons}]. In the
different scenarios discussed in the following subsections, the
$\epsilon_i$ terms will have different magnitudes, depending on the
right-handed neutrino spectrum. In those cases with significant
values of $\epsilon_i$ (i.e.~eV-scale sterile neutrinos) one must
take into account both NLO seesaw corrections and higher-order
corrections, whereas in cases with negligible $\epsilon_i$ (heavy
sterile neutrinos) one need only worry about the higher-order
correction terms, i.e.~those controlled by $y'_2$, $y'_3$, $w'_1$, $y'_\mu$ and
$y'_\tau$.

Even if $y'_2$, $y'_3$ and $w'_1$ are small and mixing corrections from the neutrino
sector are negligible, there are still effects from the charged
lepton sector. Indeed, in order to keep the solar mixing angle
within its allowed range~\cite{Schwetz:2011zk}, one has the
constraint (assuming for definiteness $y'_2=y'_3=w'_1=0$ and $\epsilon_{1,2}
\simeq 0$)
\begin{equation}
  -0.4 \leq \left(\frac{y'_\mu}{y_\mu} + \frac{y'_\tau}{y_\tau}\right) \leq 0.95\, ,
\end{equation}
on the charged lepton Yukawa couplings; the extreme choice
$y'_\mu/y_\mu = -y'_\tau/y_\tau$ gives the reactor mixing angle
\begin{equation} \label{eq:theta13max}
 \ssre \simeq 2r_1^2\left(\frac{y'_\tau}{y_\tau}\right)^2 \simeq 0.02,
\end{equation}
in both mass orderings, assuming that $y'_\tau \approx y_\tau$. In
this case $\ssatm \simeq 0.6$, and $\sssol$ retains its TBM value.

\subsection{Explicit seesaw model scenarios} \label{subsect:ss_scenarios}

In order to illustrate the versatility of the model discussed, we
present three scenarios with different mass spectra in the
right-handed neutrino sector. Each case differs by the choice of FN
charges $F_2$ and $F_3$, what one could call the ``theoretical
input''; the consequent neutrino phenomenology is described in
detail. Table~\ref{table:seesaw_summary} summarizes the key
differences in each case.

In all cases we have checked that Yukawa couplings of order 1 or 0.1
can fit the model to the active neutrino mass-squared differences
\cite{Schwetz:2011zk}, and, where appropriate, to sterile mass
parameters \cite{Kopp:2011qd}. The effects of the higher-order
corrections discussed in Sect.~\ref{subsect:high_order_ss} are
described for each scenario. Due to the large number of parameters
we will always have enough freedom to fit the masses to the data, so
that we only need to take care that mixing corrections are under
control, particularly in the IO, as discussed above.

\begin{table}[tp]
\centering \caption{Summary of the different scenarios discussed in
the $A_4$ seesaw model. In each case the WDM sterile neutrino has a
mass $M_{1} = {\cal O}({\rm keV})$, and the corresponding active
neutrino is approximately massless. } \label{table:seesaw_summary}
\vspace{8pt}
\begin{tabular*}{0.98\textwidth}{@{\extracolsep{\fill}}c|cccccc|c}
\hline \hline \T \B \multirow{2}{*}{} & \multirow{2}{*}{$F_1$,
$F_2$, $F_3$} & \multirow{2}{*}{Mass spectrum} &
\multirow{2}{*}{$|U_{\alpha 4}|$} & \multirow{2}{*}{$|U_{\alpha
5}|$} & \multicolumn{2}{c|}{$\mee$} & \multirow{2}{*}{Phenomenology}
\\ & & & & & NO & IO & \\ \hline \T I & $9$, $10$, $10$ & $M_{2,3} =
{\cal O}({\rm eV})$, & ${\cal O}(0.1)$ & ${\cal O}(0.1)$ & 0 & 0 &
$3+2$ mixing
\\[1mm] \hline
\T \multirow{2}{*}{IIA} &  \multirow{2}{*}{$9$, $10$, $0$} & $M_{2}
= {\cal O}({\rm eV}) $ &  \multirow{2}{*}{${\cal O}(0.1)$} &
\multirow{2}{*}{${\cal O}(10^{-11})$} & \multirow{2}{*}{0} &
\multirow{2}{*}{$\dfrac{2\sqrt{\dma}}{3}$ %$0.032$
}  & \multirow{5}{*}{$3+1$
 mixing}\\[1mm] && $M_{3} = {\cal O}(10^{11}\,{\rm
GeV})$ &&&& \\[1mm] \cline{1-7} \T \multirow{2}{*}{IIB} &
\multirow{2}{*}{$9$, $0$, $10$} & $M_{2} = {\cal
O}(10^{11}\,{\rm GeV})$ & \multirow{2}{*}{${\cal O}(10^{-11})$} &
\multirow{2}{*}{${\cal O}(0.1)$} & \multirow{2}{*}{$\dfrac{\sqrt{\dms}}{3}$%0.0029$
} &
\multirow{2}{*}{$\dfrac{\sqrt{\dma}}{3}$%$0.016$
} &  \\[1mm] && $M_{3} = {\cal O}({\rm eV})$ &&&&  \\[1mm] \hline \Tbig III & $9$, $5$, $5$ & $M_{2,3} = {\cal O}(10\,{\rm GeV})$ & ${\cal O}(10^{-6})$ & ${\cal O}(10^{-6})$ & $\dfrac{\sqrt{\dms}}{3}$
& $\sqrt{\dma}$%$\ls 0.049$
& Leptogenesis\\[3mm]
\hline \hline
\end{tabular*}
\end{table}

\subsubsection{\bf \it Scenario I: two eV-scale right-handed neutrinos} \label{subsubsect:I_2eV}

In this case we assign the FN charges $F_1 = 9$, $F_{2,3} = 10$, so
that the right-handed neutrino masses are lowered down to the eV
scale. It is now notable that $\epsilon_{1,2} = {\cal O}(0.1) $ can
be expected, indicating that NLO seesaw terms should be considered.
The effects are more pronounced in the IO case, since two of the
active neutrinos are nearly degenerate and are more sensitive to
corrections. The five neutrino mass eigenvalues are given by the
full expressions in Eqs.~\eqref{eq:numasses_ho} and
\eqref{eq:numasses_ho_ih}.

In this scenario, there are no heavy right-handed neutrinos that
could be used to explain the matter-antimatter asymmetry via
leptogenesis. Neutrinoless double beta decay is also vanishing since
the contributions from active and sterile neutrinos exactly cancel
each other\footnote{This is different to the usual analysis,
e.g.~in Ref.~\cite{Barry:2011wb} (see also \cite{Li:2011ss}), in which
sterile singlet states are simply added to an existing model.},
unless there are other new physics contributions. However, the
eV-scale right-handed neutrinos offer an explanation for the
short-baseline oscillation anomalies often attributed to them.

\begin{figure}[tp]
\centering \vspace{-0.cm}
\includegraphics[width=0.6\textwidth]{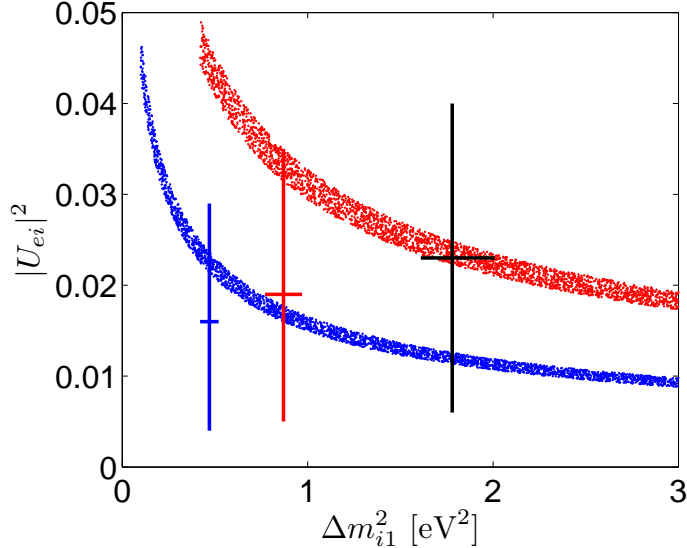}
\vspace{-0.cm}
\caption{The allowed ranges of $|U_{e4}|^2-\Delta m^2_{41}$ (blue)
and $|U_{e5}|^2-\Delta m^2_{51}$ (red) in the inverted ordering,
requiring that the oscillation parameters lie in their currently
allowed 2$\sigma$ ranges. The blue and red vertical and horizontal
error bars indicate the allowed $2\sigma$ range for the $3+2$ mass and
mixing parameters from Ref.~\cite{Kopp:2011qd}, their intersection is
the best-fit point. The black errors bars are for the $3+1$ case from
Ref.~\cite{Kopp:2011qd}, to be discussed in scenario II in
Sect.~\ref{subsubsect:II_1eV_1GeV}.  \label{fig:mass_mix}}
\end{figure}

In the NO case, one of the two sterile neutrinos could mix with
$\nu_e$ via $U_{e4} \simeq \epsilon_1$. The reactor flux loss is
therefore explained since part of the total flux of
$\overline{\nu}_e$ oscillates into sterile neutrinos. However, one
finds that the active-sterile mixing turns out to be too tiny to
account for the reactor anomaly. This can be deduced from
Eqs.~\eqref{eq:numassesb} and \eqref{eq:numassesb-inv}: at leading
order, $\epsilon^2_1 \propto m_2/m_4$. In the NO, $m_2 \simeq
0.009~{\rm eV}$ is fixed by the neutrino mass-squared differences,
and hence, $\epsilon_1$ can hardly be sizable for an eV-scale $m_4$.
The situation is different for the IO case, since $m_1\simeq m_2
\simeq 0.05~{\rm eV}$ is fixed from neutrino oscillation
experiments. Furthermore, both $U_{e4}$ and $U_{e5}$ are
non-vanishing (see Fig.~\ref{fig:mass_mix}).

The effect of higher-order operators on the active-sterile mixing is
very small. Switching on $w'_1$ gives $|U_{e5}|^2 \simeq
{\cal O}(r_1^2)\,\epsilon_2^2$ in the NO [cf.~Eq.~\eqref{eq:nlo_final_nh}],
which will still not give sufficient mixing to explain the data. In
the IO case, $|U_{e5}|^2 \simeq 4[1+{\cal O}(r_1)]\epsilon_2^2$, so the small
correction term makes little difference. Indeed, the allowed ranges
illustrated in Fig.~\ref{fig:mass_mix} already include the effects
of higher-order operators. One observes that the desired
active-sterile mixing can indeed be achieved in the IO case.

In what regards active neutrino mixing, deviations from TBM come
from both NLO seesaw terms ($\propto \epsilon_i$) and higher-order
operators ($\propto y'_2,y'_3,w'_1,y'_\mu,y'_\tau$). If we only consider
higher-order corrections in the neutrino sector for simplicity, i.e.~the $y'_{2,3}$ and $w_1'$
terms in Eqs.~\eqref{eq:nlo_md_terms} and \eqref{eq:mr_nlo_terms} respectively, then from
Eqs.~\eqref{eq:nlo_final_nh} and \eqref{eq:nlo_final_ih} only
$U_{\mu 3} \propto \epsilon_2$ receives visible corrections in the
NO, since $\epsilon_1$ and the product $\sigma^N_+ R$ are both small.
However, the higher-order terms related to the product $\sigma^I_+ G$ lead
to sizable corrections to $|U_{e 2}|^2$ in the IO case; $|U_{e 2}|$
could be enhanced or reduced depending on the signs and magnitudes of
$y'_{2,3}$ and $w_1'$. In addition, non-zero $\theta_{13}$ can be obtained from the
charged lepton corrections, as discussed in
Sect.~\ref{subsect:high_order_ss} above.

\subsubsection{\bf \it Scenario II: split seesaw with both eV-scale and heavy right-handed neutrinos} \label{subsubsect:II_1eV_1GeV}

We have shown that it is possible to get either normal or inverted
ordering by choosing the alignment of the flavon VEV $\langle
\varphi'' \rangle$ correctly [cf.~Eqs.~\eqref{eq:nh_align} and
\eqref{eq:ih_align}]. In this case we now assign different FN
charges to the two seesaw right-handed neutrinos, so that there are
four distinct possibilities, depending on the mass ordering of
active neutrinos and which sterile neutrino ($\nu_2^c$ or $\nu_3^c$)
is chosen as the heavy one. One can then use a two-stage seesaw, by
integrating out the heavy sterile neutrino first and then applying
the seesaw formula again. With the assignments $F_1 = 9$, $F_2 =
10\, (0)$ and $F_3 = 0\, (10)$ the sterile neutrino $\nu_3^c$
($\nu_2^c$) has a mass in the $10^{11}$ GeV range, and is integrated
out first, whereas $\nu_2^c$ ($\nu_3^c$) is at the eV scale. The
third (second) column of $M_D$ is then used in the seesaw formula,
leading to a $3\times3$ effective neutrino mass matrix of rank 1
that gives one of the active neutrinos masses. The full $4\times4$
mass matrix in the basis $(\nu_e,\nu_\mu,\nu_\tau,\nu^c_{2(3)})$
leads to mixing between the active sector and the remaining eV-scale
sterile neutrino $\nu^c_2$ ($\nu^c_3$). Here one can apply the
method and formulae outlined in Sect.~\ref{subsect:seesaw_analyt},
except that one has a $4\times4$ mixing matrix, which can simply be
obtained from the formulae in Eqs.~\eqref{eq:U-nor} and
\eqref{eq:U-inv} by removing the relevant row and column.

\begin{itemize}
\item {\bf Case IIA: $\nu_3^c$ heavy, ($F_3$ = 0), $\nu^c_2$ light ($F_2$ = 10)} \\
In this case one removes the fifth row and fifth column of $U_\nu$
in Eqs.~\eqref{eq:U-nor} and \eqref{eq:U-inv}, giving the same
$4\times4$ mixing matrix in both mass orderings, and the matrix
elements $|U_{e 5}|^2$ and $|U_{\mu 5}|^2$ are zero. The light
neutrino mass eigenvalues $m_i$ ($i=1,2,3,4$) are given by the
expressions in Eqs.~\eqref{eq:numasses_ho} and
\eqref{eq:numasses_ho_ih} with $\epsilon_2$ set to zero; the heavy
neutrino has the mass $M_3 = w_3u''$. It is the small value of $F_3$
that leads to $\epsilon_2 \simeq 0$ [Eq.~\eqref{eq:epsilons}], so
that $m_3$ (or $m_1$) does not receive any higher-order corrections,
as this mass originates from the high-scale of $M_3$, whose FN
charge ``cancelled'' in the leading order seesaw formula. Although
in our example we have $F_3 = 0$, so that $M_3 \simeq 10^{11}$ GeV,
even with $F_3 = 5$ and $M_3 \simeq 10$ GeV, one has $\epsilon_2
\simeq 10^{-6}$ (see scenario III), so that NLO corrections would
still be under control.

The FN charge $F_2=10$ of the eV-scale neutrino gives corrections to
$m_2$ and $m_4$, via $\epsilon_1 \simeq 0.1$. With order one Yukawas
and values for the VEVs as before, $M_3$ lies around $10^{11}$~GeV.
The effective mass in $\obb$ is given by the $(1,1)$ element of the
$4\times4$ mass matrix, which, at leading order, is
\begin{gather}
\mee^{({\rm NO})} = 0\, , \quad  \mee^{({\rm IO})} = \left|\frac{2m^{(0)}_1}{3}\right| = \frac{2\sqrt{\dma}}{3} \simeq 0.032 ~ {\rm eV} \, .
\end{gather}
Here one can see that the contribution of the light neutrino of mass
$m^{(0)}_2$ has cancelled with that of the light sterile neutrino
$\nu^c_2$, in both mass orderings. Note again that this is different
from the usually discussed effects of sterile neutrinos in $\obb$.
The effective mass is zero in the NO since at leading order,
$U_{e3}=0$. A non-zero value of $U_{e3}$ would give a very small
contribution to the effective mass in the NO, and a completely
negligible one in the IO.

\item {\bf Case IIB: $\nu_2^c$ heavy ($F_2$ = 0), $\nu^c_3$ light ($F_3$ = 10)} \\
Here the mixing matrix is found by removing the fourth row and
column of Eqs.~\eqref{eq:U-nor} and \eqref{eq:U-inv}, so that the
matrix elements in Eqs.~\eqref{eq:nlo_final_nh} and
\eqref{eq:nlo_final_ih} can be relabelled $|U_{e 5}|^2 \rightarrow
|U_{e 4}|^2$ and $|U_{\mu 5}|^2 \rightarrow |U_{\mu 4}|^2$. The
light neutrino mass eigenvalues $m_i$~($i=1,2,3,4$) are now found by
setting $\epsilon_1$ to zero in Eqs.~\eqref{eq:numasses_ho} and
\eqref{eq:numasses_ho_ih}, with the relabelling $m_5\rightarrow m_4$;
the heavy neutrino has the mass $M_2 = w_2 u'$. The roles of the
sterile neutrinos are now swapped, and $M_2$ is situated at the
$10^{11}$~GeV scale. The effective mass at leading order is
\begin{eqnarray}
\mee^{({\rm NO})} &= &\left|\frac{m^{(0)}_2}{3}\right| = \frac{\sqrt{\dms}}{3}
\simeq 0.0029~{\rm eV} \; ,\nonumber \\[1mm] \mee^{({\rm IO})} &=&
\left|\frac{m^{(0)}_2}{3}\right| \simeq \frac{\sqrt{\dma}}{3} \simeq
0.016~{\rm eV} \; ;
\end{eqnarray}
in this case the contribution of $m^{(0)}_3$ has cancelled. Again,
corrections to the mixing angles give very small corrections to the
effective mass.
\end{itemize}

In both cases IIA and IIB one could potentially explain the reactor
anomaly in the framework of $3+1$ neutrino mixing
\cite{Kopp:2011qd,Giunti:2011sn}, with $|U_{e4}|^2 \simeq [1+{\cal O}(r_1)]\,
\epsilon_1^2$ in case IIA and $|U_{e4}|^2 \simeq
4[1+{\cal O}(r_1)]\epsilon_2^2$ in the IO in case B. Once again only the IO
fits the data: the allowed ranges in the mass-mixing plane for the
IO in case IIA (IIB) are shown by the blue (red) points in
Fig.~\ref{fig:mass_mix}. One can see that the best-fit point (the
black cross) from Ref.~\cite{Kopp:2011qd} is compatible with case
IIB. Finally, the effects of higher-order operators on both
active-sterile mixing and active mixing are the same as in scenario
I, except that one should switch off the effect of $\epsilon_2$
($\epsilon_1$) in case IIA (IIB).

\subsubsection{\bf \it Scenario III: two heavy right-handed neutrinos} % (the $\nu$MSM)}
\label{subsubsect:III_2GeV}

In this case we take $F_1=9$, $F_{2,3}=5$, so that one can estimate
that the $\epsilon_i \simeq 10^{-6}$ ($i=1,2$) are dramatically
suppressed, and the NLO seesaw terms in Eqs.~\eqref{eq:U-nor} and
\eqref{eq:U-inv} can be safely neglected. The $3\times3$ effective
neutrino mass matrix is given by Eq.~\eqref{eq:seesaw}, with $M_D$
defined in Eqs.~\eqref{eq:md_nh} or \eqref{eq:md_ih} and $M_R$ from
Eq.~\eqref{eq:mr_2by2}; the active neutrino masses are simply given
by the leading order masses $m_i^{(0)}$. The heavy neutrinos have
masses $M_2=w_2u'\lambda^{10}$ and $M_3=w_3u''\lambda^{10}$. Without
the effect of the $\epsilon_i$ terms, the only modifications to the
TBM pattern come from the higher-order operators in
Sect.~\ref{subsect:high_order_ss}.

The two heavy right-handed neutrinos that participate in the seesaw
formula have masses around 5 GeV, assuming order one Yukawas and the
usual values of the VEVs. Note that one could set $w_2 = w_3$ to
obtain degenerate right-handed neutrinos $M_2 = M_3$. The choice of
degenerate sterile neutrinos in the few GeV regime would correspond
to the $\nu$MSM paradigm, in which no new scales between the SM
and the Planck scale are assumed. Baryogenesis then proceeds via
oscillations between $\nu_2^c$ and $\nu_3^c$, which need to be
sufficiently degenerate $\left(|M_2-M_3|/M_2 \simeq 10^{-6}\right)$
to give the correct baryon asymmetry \cite{Canetti:2010aw}.

If we choose $F_{2,3} = 0$ instead, then $M_{2,3} \simeq 10^{11}$ GeV,
so that the CP-violating decay of right-handed neutrinos could explain
the matter dominated Universe via thermal leptogenesis. The required
CP violation may originate from complex Yukawa couplings. We further
note that, similar to the ordinary type~I seesaw, neutrinoless double
beta decay is allowed, and the right-handed neutrinos play no role in
this process since their contribution $\sum_{i=2,3}\theta_{i}^2/M_{i}$
is strongly suppressed by the inverse of their mass. Explicitly, at
leading order the effective mass from the $(1,1)$ entry of
Eq.~\eqref{eq:seesaw} is
\begin{align}
 \mee^{({\rm NO})} &= \left|\frac{m^{(0)}_2}{3}\right| = \frac{\sqrt{\dms}}{3} \simeq 0.0029~ {\rm eV} \; , \\[1mm]
 \mee^{({\rm IO})} &= \left|\frac{2m^{(0)}_1}{3} + \frac{m^{(0)}_2}{3}\right| \simeq \sqrt{\dma} \simeq 0.049 ~{\rm eV} \; ,
\end{align}
where the mass eigenvalues are real. If $m^{(0)}_1$ and $m^{(0)}_2$ are complex, the IO case becomes $\mee^{({\rm IO})} \ls \sqrt{\dma}$. Corrections from higher order terms are again small.\\

\section{An effective theory approach} \label{sect:efftheory}

In this section we recast the idea presented in
Ref.~\cite{Barry:2011wb}, this time in the context of keV sterile
neutrino WDM rather than eV-scale sterile neutrinos. A popular
flavor symmetry model, which predicts TBM and is based on the group
$A_4$, is modified in order to accommodate a keV sterile neutrino.
Unlike the seesaw model, neutrinos get mass from effective operators
and only one sterile state is introduced. We also extend the
discussion to include the effects of higher-order operators.

\subsection{$A_4$ symmetry with one keV sterile neutrino}

The Altarelli-Feruglio (AF) $A_4$ neutrino mass
model~\cite{Altarelli:2005yp} is well known, and at leading order
gives exact TBM for the lepton flavor mixing matrix. The original AF
model includes three sets of flavon fields $\varphi$, $\varphi'$ and
$\xi$ in addition to the SM particle content. We add an additional
sterile neutrino transforming as a singlet under $A_4$ and $Z_3$,
with the $U(1)_{\rm FN}$ charge of $F_s = 8$. The relevant particle
assignments are summarized in Table~\ref{table:afmodel}.
\begin{table}[tp]
\centering \caption{Particle assignments of the $A_4$ model, modified from Ref.~\cite{Altarelli:2005yp} to include a sterile
neutrino $\nu_s$. The additional $Z_3$ symmetry decouples the charged lepton and neutrino sectors; the $U(1)_{\rm FN}$ charge
generates the hierarchy of charged lepton masses and regulates the scale of the sterile state.} \label{table:afmodel} \vspace{8pt}
\begin{tabular}{c|ccccc|cccc|c}
\hline \hline \T \B Field & $L$ & $e^c$ & $\mu^c$ & $\tau^c$ & $h_{u,d}$ & $\varphi$ & $\varphi'$ & $\xi$ & $\Theta$ & $\nu_s$ \\
\hline \T $SU(2)_L$ & $2$ & $1$ & $1$ & $1$ & $2$ & $1$ & $1$ & $1$ & $1$ & $1$ \\
$A_4$ & $\ul{3}$ & $\ul{1}$ & $\ul{1}''$ & $\ul{1}'$ & $\ul{1}$ & $\ul{3}$ & $\ul{3}$ & $\ul{1}$ & $\ul{1}$ & $\ul{1}$  \\
$Z_3$ & $\omega$ & $\omega^2$ & $\omega^2$ & $\omega^2$ & $1$ & $1$ & $\omega$ & $\omega$ & $1$ & $1$ \\
$U(1)_{\rm FN}$ & - & $4$ & $2$ & $0$ & - & - & - & - & $-1$ & $8$  \\[1mm] \hline \hline
\end{tabular}
\end{table}

As discussed in Ref.~\cite{Barry:2011wb}, at leading order the Yukawa couplings for the lepton sector read
\begin{eqnarray}\label{eq:L}
-{\cal L}_{\rm Y} &=& \frac{y_e}{\Lambda} \lambda^4\left(\varphi L h_d\right)
e^c + \frac{y_\mu}{\Lambda} \lambda^2\left(\varphi L h_d\right)' \mu^c +
\frac{y_\mu}{\Lambda} \left(\varphi L h_d\right)'' \tau^c +
\frac{x_a}{\Lambda^2}\xi(Lh_uLh_u) +
\frac{x_d}{\Lambda^2}(\varphi'Lh_uLh_u) \nonumber \\
 && +
\frac{x_e}{\Lambda^2}\lambda^8\xi(\varphi' Lh_u)\nu_s +
\frac{x_f}{\Lambda^2}\lambda^8(\varphi'\varphi'Lh_u)\nu_s + m_s\lambda^{16}\nu^c_s\nu^c_s
+ {\rm h.c.},
\end{eqnarray}
where $m_s$ is a bare Majorana mass. Note that the $A_4$ invariant
dimension-5 operator $\frac{1}{\Lambda}\lambda^8(\varphi' L
h_u)\nu_s$ is not invariant under the $Z_3$ symmetry. With the
following vacuum alignments (as in the AF model)
\begin{equation}
\langle \varphi \rangle = (v,0,0) \; , \quad \langle \varphi'
\rangle = (v',v',v') \; , \quad \langle \xi \rangle = u \; , \quad
\langle h_{u,d} \rangle =v_{u,d} \; ,
\end{equation}
the charged lepton mass matrix is diagonal
[cf.~Eq.~\eqref{eq:m_ell}], and the full $4\times 4$ neutrino mass
matrix is
\begin{equation}
M^{4\times4}_\nu = \begin{pmatrix} a+\frac{2d}{3} & -\frac{d}{3} &
-\frac{d}{3} & e \\ \cdot & \frac{2d}{3} & a-\frac{d}{3} & e \\
\cdot & \cdot & \frac{2d}{3} & e \\ \cdot & \cdot & \cdot & m_s
\end{pmatrix}, \label{eq:m4by4}
\end{equation}
where $a = 2x_a\frac{u v_u^2}{\Lambda^2}$, $d =
2x_d\frac{v'v_u^2}{\Lambda^2}$ and $e = \sqrt{2}x_e\lambda^8 \frac{u
v'v_u}{\Lambda^2}$ have dimensions of mass. The first three elements
of the fourth row of $M^{4\times4}_\nu$ are identical because of the
VEV alignment $\langle \varphi' \rangle = (v',v',v')$, which was
necessary to generate TBM in the three-neutrino case; this alignment
combined with the $A_4$ multiplication rules causes the term
proportional to $x_f$ in Eq.~\eqref{eq:L} to vanish.

If one assumes that $a < m_s$ and expands to second order in the
small ratio $e/m_s$, the mixing matrix diagonalizing
$M^{4\times4}_\nu$ in Eq.~\eqref{eq:m4by4} is~\cite{Barry:2011wb}
\begin{equation}
U \simeq \begin{pmatrix} \frac{2}{\sqrt{6}} & \frac{1}{\sqrt{3}} & 0
& 0 \\ -\frac{1}{\sqrt{6}} & \frac{1}{\sqrt{3}} &
-\frac{1}{\sqrt{2}} & 0 \\ -\frac{1}{\sqrt{6}} & \frac{1}{\sqrt{3}}
& \frac{1}{\sqrt{2}} & 0 \\ 0 & 0 & 0 & 1 \end{pmatrix} +
\begin{pmatrix} 0 & 0 & 0 & \frac{e}{m_s} \\ 0 & 0 & 0 &
\frac{e}{m_s} \\ 0 & 0 & 0 & \frac{e}{m_s} \\ 0 &
-\frac{\sqrt{3}e}{m_s} & 0 & 0 \end{pmatrix} + {\cal
O}\left(\frac{e^2}{m_s^2}\right) , \label{eq:v4}
\end{equation}
with the eigenvalues
\be
 m_1 = a+d\, ,~~  m_2 = a - \frac{3e^2}{m_s}\, ,
~~  m_3 = -a+d\, ,  ~~m_4 = m_s + \frac{3e^2}{m_s}\, .
\label{eq:nu_mass_evs} \ee As we will see, the chosen FN charge
forces $m_s$ to be at the desired keV scale and sets the magnitude
of active-sterile mixing, $e/m_s = {\cal O}(10^{-4})$. This means
that the ``seesaw contribution'' ($\propto e^2/m_s$) to $m_2$ in
Eq.~\eqref{eq:nu_mass_evs} is negligible.

\subsection{Estimation of the mass scales and active-sterile mixing}

In order to examine the viability of the model we provide a rough
numerical example. As discussed in the original AF model
\cite{Altarelli:2005yp}, we assume that $(i)$ the Yukawa couplings
$y$ and $x$ remain in a perturbative regime; $(ii)$ the flavon VEVs
are smaller than the cut-off scale and $(iii)$ all flavon VEVs fall
in approximately the same range, and obtain the following relation
constraining the flavon VEVs:
\begin{equation}
0.004 < \frac{u}{\Lambda} \approx \frac{v'}{\Lambda} \approx
\frac{v}{\Lambda} < 1 \, , \label{eq:flavonscales}
\end{equation}
with the cut-off scale $\Lambda$ ranging between $10^{12}$ and
$10^{15}$ GeV. We would like to suppress the mass of the keV
neutrino, while at the same time keep its mixing small enough and
satisfy the conditions in Eq.~\eqref{eq:flavonscales}. By
choosing the FN charge of $\nu_s$ (i.e.~$F_s = 8$) and the mass
scales
\begin{gather}
u \simeq v' \simeq 10^{10}~{\rm GeV} \; , \quad v \simeq  10^{11}~{\rm
GeV} \; , \quad \Lambda \simeq 10^{12}~{\rm GeV} \; , \notag \\ v_{u,d}
\simeq 10^2~{\rm GeV} \; , \quad \langle \Theta \rangle \simeq
10^{11}~{\rm GeV} \; ,
\end{gather}
which means that $\lambda = \langle\Theta\rangle/\Lambda \simeq 0.1$, one obtains \bea \D a \simeq d
\simeq 0.1 \left(\frac{u}{10^{10}~{\rm
GeV}}\right)\left(\frac{v_u}{10^{2}~{\rm GeV}}\right)^2
\left(\frac{10^{12}~{\rm GeV}}{\Lambda}\right)^2 \ {\rm eV}\, ,
\\[2mm] \D
e \simeq
0.1\left(\frac{\lambda}{10^{-1}}\right)^8\left(\frac{u}{10^{10}~{\rm
GeV}}\right)\left(\frac{v'}{10^{10}~{\rm
GeV}}\right)\left(\frac{v_u}{10^{2}~{\rm GeV}}\right)
\left(\frac{10^{12}~{\rm GeV}}{\Lambda}\right)^2 \ {\rm eV}\, , \eea
with the assumption that the Yukawa couplings $x_{a,d,e}$ are of
order 1.

The Majorana mass term $m_s\nu^c_s\nu^c_s$ is doubly suppressed by
the $U(1)_{\rm FN}$ charge. There are additional terms that can
give a contribution to this mass in addition to the bare term. From
the particle assignments in Table~\ref{table:afmodel}, the leading
order contribution to $m_s$ reads
\begin{eqnarray}
\left(\frac{x_s}{\Lambda}\varphi\varphi\right)\lambda^{16}\nu^c_s\nu^c_s
\Longrightarrow
\left({x_s}\frac{v^2}{\Lambda}\right)\lambda^{16}\, ,
\label{eq:majcontrib}
\end{eqnarray}
so that these terms are suppressed by $\lambda^{16}$, and the
resulting Majorana mass can be of order keV:
\begin{equation}
m_s \simeq \left(\frac{\lambda}{10^{-1}} \right)^{16}
\left(\frac{v}{10^{11}~{\rm GeV}}\right)^2 \left(\frac{10^{12}~{\rm
GeV}}{\Lambda}\right) \ {\rm keV}\, .
\end{equation}%
The active-sterile mixing is given by
\begin{equation}
 \theta_s = \frac{e}{m_s} \simeq 10^{-4}\, ,
\end{equation}
corresponding to $\sin^2\!\theta_s \simeq 10^{-8}$, in accordance
with the astrophysical constraints discussed in
Sect.~\ref{sect:gen}. It should also be noticed that in this model
the charged lepton masses are
\begin{equation}
m_\alpha = y_\alpha v_d \frac{v}{\Lambda} \lambda^{F_\alpha} \simeq 10
\left(\frac{v_d}{10^{2}~{\rm GeV}}\right)\left(\frac{v}{10^{11}~{\rm
GeV}}\right)\left(\frac{10^{12}~{\rm
GeV}}{\Lambda}\right)\left(\frac{\lambda}{10^{-1}}\right)^{F_\alpha}
\ {\rm GeV}\, ,
\end{equation}
so that we get the correct mass spectrum with the FN charges ($F_\alpha$) of
4, 2 and 0 for $e^c$, $\mu^c$ and $\tau^c$, respectively [assuming
$y_\alpha \ls {\cal O}(1)$].

\subsection{Higher-order corrections and non-zero $\theta_{13}$} \label{subsect:nlo_eff}

One may also wonder if higher-order terms could lead to significant
corrections to the lepton flavor mixing and neutrino masses so as to
generate a non-zero $\theta_{13}$, as suggested by the T2K
experiment. In general, both the neutrino and charged lepton mass
matrices receive higher-order corrections, suppressed by additional
powers of the cutoff scale $\Lambda$; those are the only type of
corrections that we consider here.

In the charged lepton sector, the NLO corrections to $M_\ell$ come
from terms like
\begin{eqnarray}
\frac{1}{\Lambda^2} \left[ y'_{e}\lambda^4 \left(\varphi \varphi L h_d\right)
e^c + y'_\mu \lambda^2\left(\varphi \varphi L h_d\right)' \mu^c  + y'_\tau
\left(\varphi \varphi L h_d\right)'' \tau^c  \right] ,
\label{eq:lep_nlo}
\end{eqnarray}
which however replicate the leading order patterns, as in the seesaw
model (see Appendix~\ref{subsect:lep_app}). The NLO corrections to
$M_\ell$ can thus be simply absorbed into the coefficients
$y_\alpha$.

As for the sterile neutrino, the NLO corrections to $m_s$ are given
by
\begin{eqnarray}
\left(\frac{x_{s'}}{\Lambda^2}\xi\xi\xi +
\frac{x_{s''}}{\Lambda^2}(\varphi'\varphi')\xi\right)\lambda^{16}\nu^c_s\nu^c_s
\Longrightarrow \left( x_{s'}\frac{u^3}{\Lambda^2} +
x_{s''}\frac{3v'^2 u}{\Lambda^2}\right)\lambda^{16}\, ;
\label{eq:majcontrib2}
\end{eqnarray}
in this case the contributions in Eq.~\eqref{eq:majcontrib2} are of
order $10^{-4}$~keV, and do not affect the scale of $m_s$
significantly. Note that the term
$\frac{x_{s''}}{\Lambda^2}\lambda^{16}(\varphi'\varphi'\varphi')\nu_s^c\nu_s^c$ is
in principle also allowed, but vanishes after $A_4$ symmetry
breaking, just like the $x_f$ term in Eq.~\eqref{eq:L}. NLO
corrections to the $e$ parameter come from terms like
\begin{equation}
\frac{x'_{e}}{\Lambda^3}\lambda^8\xi(\varphi'\varphi Lh_u)\nu_s\, ,
\end{equation}
which lead to
\begin{equation}
e' \simeq 0.01
\left(\frac{\lambda}{10^{-1}}\right)^8\left(\frac{u\,v'}{(10^{10}~{\rm
GeV})^2}\right)\left(\frac{v}{10^{11}~{\rm
GeV}}\right)\left(\frac{v_u}{10^{2}~{\rm GeV}}\right)
\left(\frac{10^{12}~{\rm GeV}}{\Lambda}\right)^3 \ {\rm eV}\; ,
\label{eq:2ndorder_e}
\end{equation}
indicating again that the active-sterile mixing is hardly affected.

The higher-order operators contributing to light neutrino masses are
of order $1/\Lambda^3$. There exist only three such terms that
cannot be absorbed by a redefinition of the parameters $a$ and
$b$~\cite{Altarelli:2005yx}, i.e.
\begin{equation}
\frac{x_1}{\Lambda^3}(\varphi\varphi')'(Lh_uLh_u)'' \; , \quad
\frac{x_2}{\Lambda^3}(\varphi\varphi')''(Lh_uLh_u)' \; ,\quad {\rm
and} \quad \frac{x_3}{\Lambda^3}\xi(\varphi Lh_uLh_u) \; ,
\end{equation}
so that the light neutrino mass matrix is modified to
\begin{eqnarray}
M_\nu = M^{(0)}_\nu + M^{(1)}_\nu = \begin{pmatrix} a+\frac{2d}{3} &
-\frac{d}{3} &
-\frac{d}{3} \\ \cdot & \frac{2d}{3} & a-\frac{d}{3}  \\
\cdot & \cdot & \frac{2d}{3}
\end{pmatrix} + \begin{pmatrix} \frac{2}{3} \eta_3 & \eta_2 & \eta_1
\\ \cdot & \eta_1 & -\frac{1}{3} \eta_3 \\ \cdot & \cdot & \eta_2
\end{pmatrix} ,
\end{eqnarray}
where $\eta_1 = 2 x_1 \frac{v v'v^2_u}{\Lambda^3}$, $\eta_2 = 2 x_2
\frac{v v'v^2_u}{\Lambda^3}$ and $\eta_3 = 2 x_3 \frac{u v
v^2_u}{\Lambda^3}$. For ${\cal O}(1)$ Yukawa couplings, one can estimate
that
\begin{equation}
\eta_i \simeq 0.01 \left(\frac{v}{10^{11}~{\rm GeV}}\right)
\left(\frac{v'}{10^{10}~{\rm
GeV}}\right)\left(\frac{v_u}{10^{2}~{\rm GeV}}\right)^2
\left(\frac{10^{12}~{\rm GeV}}{\Lambda}\right)^3 \ {\rm eV}\, .
\label{eq:2ndorder_ad}
\end{equation}
As a result, the NLO terms may lead to visible modifications to the
TBM pattern, in particular to $\theta_{13}$, but on the other
hand do not entirely spoil the leading order picture, since one always has enough parameters to fit the data. Keeping only first order terms in
$\eta_i$, one obtains
\begin{eqnarray} \nonumber
m_1 & \simeq & a + b - \frac{1}{2} \left( \eta_1+\eta_2 \right) +
\frac{1}{3} \eta_3 \; , \\
m_2 & \simeq & a + \eta_1+\eta_2 \; , \\
m_3 & \simeq & -a + b + \frac{1}{2} \left( \eta_1+\eta_2 \right) +
\frac{1}{3} \eta_3 \; ,\nonumber
\end{eqnarray}
together with the mixing angles
\begin{eqnarray}\nonumber
\ssre & \simeq & \frac{(\eta_1
-\eta_2)^2}{8 a^2} \; , \\[1mm]
\sssol & \simeq & \frac{1}{3}\left(1 - \frac{2\eta_3}{3b}\right)  ,\\[1mm]
\ssatm & \simeq & \frac{1}{2}\left(1 - \frac{\eta_1 -\eta_2}{4a}\right)  .\nonumber
\end{eqnarray}
As one numerical example, we take $\eta_2=-\eta_1= 0.1 a $ and
$\eta_3=0.1 b$, and obtain $\sin^2\theta_{13} \simeq 0.005$, which
is compatible with the current global-fit data at $2\sigma$ C.L. In
addition, $\sin^2\theta_{23} \simeq 0.53$ and $\sin^2\theta_{12}
\simeq 0.31$ are predicted, in good agreement with their best-fit
values~\cite{Schwetz:2011qt,Schwetz:2011zk}.

\section{Conclusion} \label{sect:summary}

The addition of sterile right-handed neutrinos to the SM is a natural
way to explain light active neutrino masses via the seesaw mechanism.
This works even if the scale of the sterile neutrinos is not equal to
its ``natural value'' of $10^{10}$ to $10^{15}$ GeV, as long as the Dirac mass
matrix can also be suppressed such that $M_D^2 /M_R$ is small.
At the same time, several observations point to sterile neutrinos at
the keV and eV scales. Therefore we have attempted, as a proof of
principle, to construct a seesaw model for neutrino mass and lepton
mixing that can provide a common framework for all these issues.

Starting from a flavor symmetry model based on the
tetrahedral group $A_4$, we described different ways to introduce
sterile neutrinos, using the seesaw mechanism (and also an
effective theory approach). In both cases the Froggatt-Nielsen (FN)
mechanism is used to suppress the masses of the right-handed
neutrinos. We stress that its presence in flavor symmetry models can be
considered necessary in order to generate the observed strong
hierarchy in the charged lepton sector. In fact, we utilize the very
same FN for both the charged lepton masses and the right-handed neutrinos.

In the seesaw model we studied different possible spectra in the
sterile sector: once the keV WDM neutrino is decoupled one can have
the remaining two neutrinos at the eV scale or at a high scale (in our
example at either 10 GeV or close to the flavor symmetry breaking
scale of $\simeq 10^{11}$ GeV). In each case there are distinct
phenomenological consequences, both for neutrino mass and neutrinoless
double beta decay. In particular, NLO corrections to the seesaw
formula need to be taken into account when the sterile neutrinos are
at the eV scale.

Motivated by the recent indications for nonzero $\theta_{13}$ in the
T2K experiment, we examined the effect of higher-order terms in
both the seesaw model and the effective theory. In general active
neutrino mixing angles will receive corrections of the same order.
We highlighted the fact that active-sterile mixing is stable
in any seesaw model, being defined as the ratio of two large scales.

Although one can explain both eV-scale and keV-scale sterile
neutrinos in a single framework, it is not possible to have viable
WDM, eV-scale neutrinos and heavy neutrinos for leptogenesis in a
model containing three right-handed neutrinos. However, we emphasize
the point that if one departs from the common theoretical prejudice
of right-handed neutrinos residing at around the Grand Unification
scale, various interesting model building options can arise. Further
experimental data in the years to come will put the presence of
sterile neutrinos at the eV and/or keV scale/s to the test, thus
determining whether it is indeed a useful enterprise to further
pursue this avenue of research.

\begin{acknowledgments}
We thank T.~Asaka and J.~Heeck for helpful discussions. This work was
supported by the ERC under the Starting Grant MANITOP and by the
Deutsche Forschungsgemeinschaft in the Transregio 27 ``Neutrinos and
beyond -- weakly interacting particles in physics, astrophysics and cosmology''.
\end{acknowledgments}

%\newpage
\renewcommand{\theequation}{A-\arabic{equation}}
  % redefine the command that creates the equation no.
\setcounter{equation}{0}
%\roman{section}

%\newpage
\begin{appendix}

\section{Corrections from higher-order operators in the seesaw model} \label{sect:highorder_app}

Here we give details of the procedure followed to calculate
corrections to the lepton mixing matrix in the presence of
higher-order operators, which affect $M_\ell$, $M_D$ and $M_R$. We only take into account corrections of relative order $r_1
\simeq 0.1$ [cf.~Eq.~\eqref{eq:VEV_ratios}]. Explicit expressions
for the corrected neutrino mass eigenvalues are also reported.

\subsection{Charged lepton sector} \label{subsect:lep_app}

The corrections to $M_\ell$ from dimension-six operators come from
coupling a second $A_4$ triplet or an $A_4$ singlet to each mass
term. The addition of the flavon $\varphi$ replicates the leading
order pattern, since the triplet from the product $(\varphi
\varphi)_3$ has a VEV in the same direction as $\varphi$
\cite{Altarelli:2005yx}. Terms with the additional singlet $\xi''$
also leave the structure of the mass matrix unchanged, but the
additional terms
\begin{equation}
 \frac{y'_{e}}{\Lambda^2}\lambda^3\xi(\varphi'Lh_d)e^c \ , \frac{y''_{e}}{\Lambda^2}\lambda^3\xi'(\varphi''Lh_d)''e^c \quad {\rm and} \quad \frac{y'''_{e}}{\Lambda^2}\lambda^3(\varphi'\varphi''Lh_d)e^c
  \label{eq:nlo_lep_terms}
\end{equation}
are also present, for all three flavors. The first term gives the largest NLO contribution, i.e.
\begin{equation}
 \delta M^{(1)}_{\ell} = \frac{v_d u v'}{\Lambda^2}\begin{pmatrix} y'_e\lambda^3 & y'_\mu\lambda & y'_\tau \\ y'_e\lambda^3 & y'_\mu\lambda & y'_\tau \\ y'_e\lambda^3 & y'_\mu\lambda & y'_\tau \end{pmatrix},
\label{eq:m_ell_nlo1}
\end{equation}
of relative order $r_1 \simeq 0.1$. The matrix diagonalizing $(M_\ell+ \delta M^{(1)}_{\ell})(M_\ell+\delta M^{(1)}_{\ell})^{\dagger}$ can be approximated by
\begin{equation}
 V_\ell \simeq \begin{pmatrix} 1 & \frac{y'_\mu}{y_\mu}r_1 & \frac{y'_\tau}{y_\tau}r_1 \\[1.5mm] -\frac{y'_\mu}{y_\mu}r_1 & 1 & \frac{y'_\tau}{y_\tau}r_1%\left(\frac{y'_\tau}{y_\tau}+\frac{y'_\mu y_\mu}{y_\tau^2}\right)
\\[1.5mm] -\frac{y'_\tau}{y_\tau}r_1 & -\frac{y'_\tau}{y_\tau}r_1%\left(\frac{y'_\tau}{y_\tau}+\frac{y'_\mu y_\mu}{y_\tau^2}\right)
& 1 \end{pmatrix} + {\cal O}(r_1^2,\lambda^2)\, ,
 \label{eq:vlep}
\end{equation}
and the charged lepton masses become
\begin{equation}
 m'_\alpha = (y_\alpha+y'_\alpha r_1) \frac{v_d v}{\Lambda} \lambda^{F_\alpha}, \qquad (\alpha = e,\mu,\tau),
\end{equation}
which amounts to a rescaling of Yukawa couplings.

\subsection{Neutrino sector} \label{sect:nu_app}

Similarly to $M_\ell$, corrections to $M_D$ from adding the singlet $\xi''$ retain the leading order form, but there are also several terms with two triplet flavons. The latter are all suppressed by $r_2 \simeq 0.01$ and can
be safely neglected. Of the nine different invariant dimension-six operators with one triplet and one singlet flavon, there are three of relative order $r_1 \simeq 0.1$, namely
\begin{equation}
  \frac{y'_1}{\Lambda^2}\lambda^{F_1}\xi(\varphi'Lh_u)\nu_1^c + \frac{y'_2}{\Lambda^2}\lambda^{F_2}\xi(\varphi''Lh_u)''\nu_2^c + \frac{y'_3}{\Lambda^2}\lambda^{F_3}\xi(\varphi Lh_u)\nu_3^c\, ,
  \label{eq:nlo_md_terms}
\end{equation}
leading to the corrections
\begin{equation}
 \delta M_D^{(1N)} = \frac{v_u u}{\Lambda^2}\begin{pmatrix} y'_1v' & -y_2'v'' & y'_3 v \\ y'_1v' & y_2'v'' & 0 \\ y'_1 v' & 0 & 0 \end{pmatrix}F \quad {\rm and} \quad \delta M_D^{(1I)} = \frac{v_u u}{\Lambda^2}\begin{pmatrix} y'_1v' & -y'_2v'' & y'_3 v \\ y'_1v' & -y_2'v'' & 0 \\ y'_1 v' & 2y'_2 v'' & 0 \end{pmatrix}F\, ,
\label{eq:md_nlo_ni}
\end{equation}
in the normal and inverted ordering, respectively. Here the matrix of FN charges is
\begin{equation}
 F={\rm diag}(\lambda^{F_1},\lambda^{F_2},\lambda^{F_3})\, .
\end{equation}

The corrections to $M_R$ come from terms with
two singlets and those with two triplets, e.g.
\begin{equation}
 \frac{w'_1}{\Lambda}\lambda^{F_1+F_3}\xi\xi\nu_1^c\nu_3^c + \ldots \quad {\rm and} \quad \frac{w''_1}{\Lambda}\lambda^{F_1+F_3}(\varphi\varphi')\nu_1^c\nu_3^c + \ldots\, ;
\label{eq:mr_nlo_terms}
\end{equation}
the singlet terms give the contribution
\begin{equation}
\delta M^{(1)}_R \propto \frac{1}{\Lambda}\begin{pmatrix}
uu''\lambda^{2F_1} & 0 & uu\lambda^{F_1+F_3} \\ \cdot &
u'u''\lambda^{2F_2} & u'u'\lambda^{F_2+F_3} \\ \cdot & \cdot &
u''u''\lambda^{2F_3} \end{pmatrix}\, , \label{eq:m_r_nlo_1}
\end{equation}
whereas the triplet terms are all suppressed by $r_2 \simeq 0.01$.
Comparison of the LO and NLO terms shows that the large ratio $r_1
\simeq 0.1$ only occurs in the $(1,3)$ element of $\delta
M^{(1)}_R$, whereas the diagonal and $(2,3)$ elements receive small
corrections of order $r_2 \simeq 0.01$. Ignoring the latter, the new
mass matrix is
\begin{equation}
M'_R = M_R+\delta M^{(1)}_R = F\begin{pmatrix} w_1 u & 0 & w'_1 u r_1 \\ \cdot & w_2 u' & 0 \\ \cdot & \cdot & w_3 u'' \end{pmatrix}F\,.
\label{eq:mr_nlo_2}
\end{equation}
% where the matrix of FN charges is
% \begin{equation}
%  F={\rm diag}(\lambda^{F_1},\lambda^{F_2},\lambda^{F_3})\, .
% \end{equation}
It is convenient to factor out the FN charges here, since they do
not appear in the leading order seesaw formula. However, as
emphasized before, they will play a role when considering NLO seesaw
terms. Expanding in the small ratios $r_1 \simeq \frac{w_3u''}{w_1u}
\simeq 0.1$, the matrix diagonalizing $M'_R$ can be approximated as
\begin{equation}
 V_R \simeq F^{-1}\begin{pmatrix} 1
& 0 & -\frac{w'_1}{w_1}r_1 \\
0 & 1 & 0 \\ \frac{w'_1}{w_1}r_1 & 0 & 1
\end{pmatrix}F + {\cal O}\left(\frac{w_3u''}{w_1u}r_1,r_1^2\right),
 \label{eq:vr}
\end{equation}
with the mass eigenvalues
\begin{align}
 M'_1 &= w_1 u \lambda^{2F_1}\left(1 + \frac{{w'_1}^2}{w_1^2}r_1^2\right)\, , \notag \\ M'_2 &= w_2 u'\lambda^{2F_2} \,  ,\label{eq:mrevs_nlo} \\ M'_3  &= w_3 u''\lambda^{2F_3}\left(1 - \frac{{w'_1}^2}{w_1^2}r_1^2\right). \notag
\end{align}
This shows that corrections to the masses $M_{1,3}$ are suppressed by $r_1^2$, and the WDM candidate $\nu_1^c$ remains in the keV range.

The diagonalization matrix in Eq.~\eqref{eq:vr} can be absorbed into $M_D$, so that the leading order neutrino mass matrix is
\begin{equation}
 M'_\nu = -M_D'{\rm diag}({M'_1}^{-1},{M'_2}^{-1},{M'_3}^{-1}){M_D'}^T\, ,
 \label{eq:mnu_nlo}
\end{equation}
where $M_D' = \left(M_D+\delta M_D^{(1)}\right) V^*_R$ and the FN charges have cancelled. The Dirac mass matrices in Eqs.~\eqref{eq:md_nh} and \eqref{eq:md_ih} plus the corrections terms in Eq.~\eqref{eq:md_nlo_ni} lead to
\begin{align}
 {M'_D}^{({\rm NO})} &= \frac{v_u}{\Lambda}\begin{pmatrix} y_1 v + y'_1v'r_1 %+y'_3v\frac{w'_1}{w_1}r_1^2 
& y_2 v'-y'_2v''r_1 & \left(y'_3v -y_1 v \frac{w'_1}{w_1}\right)r_1%-y'_1v'\frac{w'_1}{w_1}r_1^2 
\\[2mm] \left(y'_1v' -y_3 v''\frac{w'_1}{w_1}\right) r_1
 %\left(1+\frac{u''}{u}\right)
 & y_2 v'+y'_2v''r_1 & -y_3 v'' \\[2mm] \left(y'_1v'+y_3 v''\frac{w'_1}{w_1}\right) r_1
 %\left(1+\frac{u''}{u}\right)
 & y_2 v' & y_3 v''\end{pmatrix}F, \notag \\[2mm]
 {M'_D}^{({\rm IO})} & = \frac{v_u}{\Lambda} \begin{pmatrix} y_1 v + \left(y'_1v' + 2 y_3 v''\frac{w'_1}{w_1}\right) r_1 & y_2 v'- y'_2v''r_1 & 2 y_3 v''+ \left(y'_3v -y_1 v \frac{w'_1}{w_1}\right)r_1 \\[2mm]
 \left(y'_1v'-y_3 v''\frac{w'_1}{w_1} \right)r_1 & y_2 v'-y'_2v''r_1 & -y_3 v'' \\[2mm] \left(y'_1v'-y_3 v''\frac{w'_1}{w_1}\right) r_1 & y_2 v' +2y'_2v''r_1 & -y_3 v'' \end{pmatrix}F\, , \label{eq:md_nlo}
\end{align}
to first order in $r_1$, in the NO and IO, respectively. As shown
explicitly in the main text, the dynamics of the right-handed sector
are relatively unaffected: the new entries in the first column of
the Dirac mass matrices in Eq.~\eqref{eq:md_nlo} will induce mixing
between the sterile neutrino $\nu_1^c$ and the $\mu$ and $\tau$
flavors, but of the same magnitude as the original $\theta_{e1}$, so
that $\theta_{1}^2$ will not increase by that much
[cf.~Eqs.~\eqref{eq:theta_e1_nlo} and \eqref{eq:theta_mutau1_nlo}].
Thus the entire first column of $M_D'$, suppressed by the mass $M'_1
= {\cal O}({\rm keV})$, can be decoupled from the seesaw (assuming
that $|w'_1|\ls |w_1|$). In addition, corrections to $U_{e5}$ in
Eqs.~\eqref{eq:U-nor} and \eqref{eq:U-inv} will also be small (see
Sects.~\ref{subsubsect:I_2eV} and \ref{subsubsect:II_1eV_1GeV} for a
discussion of those effects).

The full $5\times5$ NLO neutrino mass matrix ${M'_\nu}^{5\times5}$
can now be constructed from the second and third columns of $M_D'$
and ${\rm diag}(M'_2,M'_3)$, as in Eq.~\eqref{eq:mnu_5by5}. Since we
consider scenarios where NLO seesaw terms are important, we once
again perform the full $5\times5$ diagonalization
[cf.~Eqs.~\eqref{eq:U-nor} and \eqref{eq:U-inv}], including the new
terms from higher-order operators in Eq.~\eqref{eq:md_nlo}. The
matrix diagonalizing ${M'_\nu}^{5\times5}$ is explicitly given by
\begin{equation}
 U_\nu = \begin{pmatrix} U_{\rm TBM} & 0_{3\times2} \\ 0_{2\times3} & 1_{2\times2} \end{pmatrix} + \delta U
\label{eq:u_nu_nlo}
\end{equation}
where, to first order in $r_1$ and second order in $\epsilon_i$,
{\small \begin{align}
 &{\delta U}^{({\rm NO})} \simeq 
%First order matrix
\begin{pmatrix} \frac{\rho_2}{\sqrt{6}} & -\frac{\rho_2}{\sqrt{3}} & -\frac{1}{\sqrt{2}}\left(\chi-\rho_3+\sigma^N_+\frac{R}{3}\right) & (1-\rho_2)\epsilon_1 & (\rho_3-\chi)\epsilon_2 \\[3.5mm]
-\frac{\sigma^N_-}{\sqrt{6}} & -\frac{1}{2\sqrt{3}}\left(\sigma^N_-+\sigma^N_+R\right) & -\frac{\sigma^N_+}{3\sqrt{2}}R & (1+\rho_2)\epsilon_1 & -\epsilon_2 \\[3.5mm] 
\frac{\sigma^N_+}{\sqrt{6}} & \frac{\sigma^N_+}{2\sqrt{3}}(1+R) &  -\frac{\sigma^N_+}{3\sqrt{2}}R & \epsilon_1 & \epsilon_2 \\[3.5mm] 
0 & -\sqrt{3}\epsilon_1 & \frac{\sigma^N_+}{\sqrt{2}}(1+R)\epsilon_1 & 0 & 0 \\[3.5mm] 0 & 
-\frac{\sigma^N_+}{\sqrt{3}}R\epsilon_2 & -\sqrt{2}\epsilon_2 & 0 & 0\end{pmatrix} \notag \\[3mm]
%Second order matrix
 &+ \begin{pmatrix} 0 & -\frac{\sqrt{3}}{2}(1-\rho_2)\epsilon_1^2 & 
\frac{1}{2\sqrt{2}}\left[2(\chi-\rho_3)\epsilon_2^2+\sigma^N_+\left(1+R\right)\epsilon_1^2\right] & 0 & 0 \\[3.5mm] 0 & -\frac{1}{2\sqrt{3}}\left[3(1+\rho_2)\epsilon_1^2 - \sigma^N_+ R \epsilon_2^2\right] & \frac{1}{2\sqrt{2}}\left[2\epsilon_2^2 + 
\sigma^N_+\left(1+R\right)\epsilon_1^2\right] & 0 & 0 \\[3.5mm] 0 & -\frac{1}{2\sqrt{3}}\left[3\epsilon_1^2 + 
\sigma^N_+ R\epsilon_2^2\right] & -\frac{1}{2\sqrt{2}}\left[2\epsilon_2^2 - \sigma^N_+\left(1+R\right)\epsilon_1^2\right] & 0 & 0 \\[3.5mm] 0 & 0 & 0 & -\frac{3}{2}\epsilon_1^2 & \frac{1}{2}\sigma^N_+\epsilon_1\epsilon_2 \\[3.5mm] 0 & 0 & 0 & 
\frac{1}{2}\sigma^N_+\epsilon_1\epsilon_2 & -\epsilon_2^2 \end{pmatrix}\, ,
\end{align}}
in the normal ordering, where only first order terms in $R\simeq{\cal O}(10^{-1})$ are kept [see Eq.~\eqref{eq:param_defs}], and $\sigma^N_{\pm} = \chi \pm \rho_2-\rho_3$. The new mass eigenvalues are
{\small \begin{align}
 m'_1 &= 0\, , \notag \\
 m'_2 &\simeq m_2^{(0)}\left\{1 -3\epsilon_1^2 -\frac{\rho_2}{3}\sigma^I_--\frac{1}{2}\left[9\rho_2^2-4\rho_2(\chi-\rho_3)-(\chi-\rho_3)^2\right]\epsilon_1^2 \right. \notag \\ &\left. \qquad \qquad -\,\frac{\sigma^N_+}{3}R\left[\rho_2(1-3\epsilon_1^2)-\sigma^N_+\epsilon_2^2\right]\right\}\, ,\notag \\[1.5mm]
 m'_3 &\simeq m_3^{(0)}\left\{1 -2\epsilon_2^2 + (\chi-\rho_3)^2(1-3\epsilon_2^2) -\frac{(\sigma^N_+)^2}{2}(1+2R)\epsilon_1^2 \right. \label{eq:numasses_ho} \\ &\left. \qquad \qquad {} +\frac{1}{6}\left[\rho_2^2+4\rho_2(\chi-\rho_3)+3(\chi-\rho_3)^2\right]R(1-2\epsilon_2^2)\right\} \notag \\[1.5mm]
 m'_4 &\simeq w_2u'\lambda^{2F_2} - m_2^{(0)}\left\{1 - 3\epsilon_1^2 +\frac{2\rho_2^2}{3}(1-6\epsilon_1^2)- 
\frac{3(\sigma^N_+)^2}{8}\frac{m_3^{(0)}}{m_2^{(0)}}\epsilon_1^2\right\}\, , \notag \\[1.5mm]
 m'_5 &\simeq w_3u''\lambda^{2F_3} -m_3^{(0)}\left\{1- 2\epsilon_2^2 +\frac{1}{2}(\chi-\rho_3)^2 - %\frac{y_1^2v^2}{4y_3^2v''^2}
\frac{1}{4} \left[8(\chi-\rho_3)^2+(\sigma^N_+)^2R\right]\epsilon_2^2\right\}\, , \notag
\end{align}}
which corresponds to Eq.~\eqref{eq:numassesb} in the limit $(\chi,\rho_2,\rho_3)
\rightarrow 0$. Here one can explicitly see that NLO seesaw
corrections are controlled by $\epsilon_i$, whereas corrections from
higher-order operators are controlled by $\chi$, $\rho_2$ and $\rho_3$. In those scenarios
where the $\epsilon_i$ are negligible, i.e.~scenario III, one could
still have corrections from the latter. Those turn out to be
small in the normal ordering.

In the inverted ordering, we have
{\small \begin{align}
\hspace{-1.0cm}&{\delta U}^{({\rm IO})} \simeq \begin{pmatrix}
\frac{1}{3\sqrt{6}}\left(3\rho_2 + \sigma^I_+ G\right) & -\frac{1}{3\sqrt{3}}\left(3\rho_2 + \sigma^I_+ G\right) & -\frac{\rho_2}{\sqrt{2}} & (1-\rho_2)\epsilon_1 & (2-\chi+\rho_3)\epsilon_2 \\[3.5mm]
\frac{1}{3\sqrt{6}}\left(3\rho_2 + \sigma^I_+ G\right) & -\frac{1}{6\sqrt{3}}\left(6\rho_2-\sigma^I_+ G\right) & -\frac{\rho_2}{\sqrt{2}} & (1-\rho_2)\epsilon_1 & -\epsilon_2 \\[3.5mm]
\frac{1}{3\sqrt{6}}\left(3\rho_2 + \sigma^I_+ G\right) & \frac{1}{6\sqrt{3}}\left(12\rho_2+\sigma^I_+ G\right) & -\frac{\rho_2}{\sqrt{2}} & (1+2\rho_2)\epsilon_1 & -\epsilon_2 \\[3.5mm]
-\frac{\sigma^I_+}{\sqrt{6}}G\,\epsilon_1 & -\sqrt{3}\,\epsilon_1 & 0 & 0 & 0 \\[3.5mm]
-\sqrt{\frac{2}{3}}\left(3 - \chi+\rho_3\right)\epsilon_2 & \frac{\sigma^I_+}{\sqrt{3}}\left(1+G\right)\,\epsilon_2 & 0 & 0 & 0 \end{pmatrix} \\[2mm]
%second order matrix
  \hspace{-1.0cm}&+ \begin{pmatrix} -\frac{1}{2\sqrt{6}}\left[2\left(6-5\left(\chi-\rho_3\right)\right)\epsilon_2^2+\sigma^I_+ G \epsilon_1^2\right] & -\frac{1}{2\sqrt{3}}\left[3(1-\rho_2)\epsilon_1^2-2\sigma^I_+(1+G)\epsilon_2^2\right] & 0 & 0 & 0 \\[3.5mm]
\frac{1}{2\sqrt{6}}\left[2(3-\chi+\rho_3)\epsilon_2^2 -\sigma^I_+ G\epsilon_1^2\right]& -\frac{1}{2\sqrt{3}}\left[3(1-\rho_2)\epsilon_1^2+\sigma^I_+(1+G)\epsilon_2^2\right] & 0 & 0 & 0 \\[3.5mm]
\frac{1}{2\sqrt{6}}\left[2(3-\chi+\rho_3)\epsilon_2^2 -\sigma^I_+ G\epsilon_1^2\right] & -\frac{1}{2\sqrt{3}}\left[3(1+2\rho_2)\epsilon_1^2+\sigma^I_+(1+G)\epsilon_2^2\right] & 0 & 0 & 0 \\[3.5mm]
 0 & 0 & 0 & -\frac{3}{2}\epsilon_1^2 & \frac{\sigma^I_+\epsilon_1\epsilon_2}{2} \\[3.5mm] 0 & 0 & 0 & \frac{\sigma^I_+\epsilon_1\epsilon_2}{2} & -\left(3-2\chi+2\rho_3\right)\epsilon_2^2\end{pmatrix}
\, , \notag
\end{align}}
to first order in $\chi$ and second order in $\epsilon_i$, where $\sigma^I_+$ and $G
= {\cal O}(10^2)$ are defined in Eq.~\eqref{eq:param_defs}. In this
case we cannot expand in $G$, in contrast to the NO case, where we
expanded to first order in $R$. The new mass eigenvalues are
{\begin{align}
 m'_1 &\simeq m_1^{(0)}\left\{1-6\epsilon_2^2-
\frac{1}{9}\left(\chi-\rho_3\right)\left(6-\sigma^I_-\right)+\left[8(\chi-\rho_3)+3\rho_2^2+4\rho_2(\chi-\rho_3)-4(\chi-\rho_3)^2\right]\epsilon_2^2\right. \notag \\[1mm] \qquad &\left. + \frac{1}{18}\left[9\rho_2^2-(\chi-\rho_3)^2\right]G  +\left[3\rho_2^2+4\rho_2(\chi-\rho_3)+(\chi-\rho_3)^2\right]G\epsilon_2^2+
\frac{(\sigma^I_+)^2}{18}G^2\left(1-3\epsilon_1^2\right)\right\}\, ,\notag
\\[1.5mm] m'_2 &\simeq m_2^{(0)}\left\{1-3(1+6\rho_2^2)\epsilon_1^2+4\rho_2^2+\frac{1}{18}\left[27\rho_2^2+12\rho_2(\chi-\rho_3)+(\chi-\rho_3)^2\right]G(1-3\epsilon_1^2) \right. \notag \\[1mm] \qquad &\left. -\frac{(\sigma^I_+)^2}{18}\left[6(1+2G)\epsilon_2^2-G^2(1-6\epsilon_2^2)\right]\right\}\, , \notag \\
m'_3 &= 0\, , \label{eq:numasses_ho_ih} \\[1.5mm]
m'_4 &\simeq w_2u'\lambda^{2F_2} - m_2^{(0)}\left\{1 +2\rho_2^2- 3(1+4\rho_2^2)\epsilon_1^2 -
\frac{(\sigma^I_+)^2}{8}\frac{m_1^{(0)}}{m_2^{(0)}}\,\epsilon_1^2\right\}\, , \notag \\[1.5mm]
m'_5 &\simeq w_3u''\lambda^{2F_3} - m_1^{(0)}\left\{1-6\epsilon_2^2-\frac{1}{6}\left[4(\chi-\rho_3)-(\chi-\rho_3)^2\right] \right. \notag \\[1mm] \qquad &\left.  +\left[8(\chi-\rho_3)-\frac{14}{3}(\chi-\rho_3)^2\right]\epsilon_2^2 -\frac{(\sigma^I_+)^2}{4}\frac{3m_2^{(0)}}{2m_1^{(0)}}\epsilon_2^2\right\} \notag\, .
\end{align}}
In this case the corrections very much depend on the scenario
concerned, since the value of the $\epsilon_i$ terms can give
cancellations. However, the correction to $|U_{e2}|^2$ constrains
the parameters $\chi$, $\rho_2$ and $\rho_3$ to be small (see discussion in the main text),
and since $G$ always occurs together with one of the three parameters the effect of $G =
{\cal O}(10^2)$ will always be suppressed. In the end we always have
enough parameters to fit the mass eigenvalues to the data.
\end{appendix}

\bibliography{bib}

\end{document}